\documentclass[a4paper,fleqn,usenatbib,useAMS]{mnras}
\pdfoutput=1
\usepackage{graphicx}	
\usepackage{amsmath}	
\usepackage{amssymb}	
\usepackage{multicol}        
\usepackage{url}
\usepackage{subfig}
\usepackage[T1]{fontenc}
\usepackage{ae,aecompl}
\usepackage{newtxtext,newtxmath}

\newcommand{\doublecolumnwidth}{480.0pt}

\defcitealias{taylor09}{TSS09}

\title[The n$\pi$-ambiguity in the NVSS RM Catalogue]{A Broadband Spectro-polarimetric View of the NVSS Rotation Measure Catalogue I: Breaking the $n\pi$-ambiguity}

\author[Y.\ K.\ Ma et al.]{Yik Ki Ma$^1$\thanks{Contact e-mail: \href{mailto:ykma@mpifr-bonn.mpg.de}{ykma@mpifr-bonn.mpg.de}}\thanks{Member of the International Max Planck Research School (IMPRS) for Astronomy and Astrophysics at the Universities of Bonn and Cologne}, 
S.~A.~Mao$^1$, 
Jeroen Stil$^2$, 
Aritra Basu$^{3,1}$, 
Jennifer West$^4$, 
Carl Heiles$^5$, 
\newauthor 
Alex S.~Hill$^{6,7,8}$, and 
S.~K.~Betti$^9$
\\
$^1$Max-Planck-Institut f\"{u}r Radioastronomie, Auf dem H\"{u}gel 69, 53121 Bonn, Germany \\
$^2$Department of Physics and Astronomy, University of Calgary, 2500 University Drive NW, Calgary, AB T2N 1N4, Canada \\
$^3$Fakult\"{a}t f\"{u}r Physik, Universit\"{a}t Bielefeld, Postfach 100131, 33501 Bielefeld, Germany \\
$^4$Dunlap Institute for Astronomy and Astrophysics, The University of Toronto, 50 St.\ George Street, Toronto, ON M5S 3H4, Canada \\
$^5$Department of Astronomy, University of California, Berkeley, CA 94720-3411, USA \\
$^6$Department of Physics and Astronomy, The University of British Columbia, Vancouver, BC V6T 1Z1, Canada \\
$^7$Space Science Institute, Boulder, CO, USA \\
$^8$National Research Council Canada, Herzberg Program in Astronomy and Astrophysics, Dominion Radio Astrophysical Observatory, PO Box 248, Penticton, \\
\phantom{$^0$}BC V2A 6J9, Canada \\
$^9$Department of Astronomy, University of Massachusetts, 710 North Pleasant Street, Amherst, MA 01003-9305, USA
}

\date{Accepted 2019 May 8. Received 2019 May 7; in original form 2018 September 7}

\pubyear{2019}

\begin{document}
\label{firstpage}
\pagerange{\pageref{firstpage}--\pageref{lastpage}}
\maketitle

\begin{abstract}
The NRAO VLA Sky Survey (NVSS) Rotation Measure (RM) catalogue is invaluable for the study of cosmic magnetism. However, the RM values reported in it can be affected by $n\pi$-ambiguity, resulting in deviations of the reported RM from the true values by multiples of $\pm 652.9\,{\rm rad\,m}^{-2}$. We therefore set off to observationally constrain the fraction of sources in the RM catalogue affected by this ambiguity. New broadband spectro-polarimetric observations were performed with the Karl G.\ Jansky Very Large Array (VLA) at $1$--$2\,{\rm GHz}$, with 23 $n\pi$-ambiguity candidates selected by their peculiarly high $|{\rm RM}|$ values. We identified nine sources with erroneous RM values due to $n\pi$-ambiguity and 11 with reliable RM values. In addition, we found two sources to be unpolarised and one source to be inconsistent with neither $n\pi$-ambiguity nor reliable RM cases. By comparing the statistical distributions of the above two main classes, we devised a measure of how much a source's RM deviates from that of its neighbours: $\Delta/\sigma$, which we found to be a good diagnostic of $n\pi$-ambiguity. With this, we estimate that there are at least 50 sources affected by $n\pi$-ambiguity among the 37,543 sources in the catalogue. Finally, we explored the Faraday complexities of our sources revealed by our broadband observations.
\end{abstract}


\section{INTRODUCTION} \label{sec:intro}
Magnetic fields are ubiquitous in the Universe. For astrophysical processes such as star formation, cosmic ray propagation, galactic outflows, and galactic evolution, magnetic fields are critical and must be considered \citep[see review by][]{beck13,beck15}. Magnetic field structures of astrophysical objects can be directly measured through their polarised synchrotron diffuse emission \citep[e.g.,][]{kothes08,heald09,ma16,basu17}. However, this technique is limited to probing volumes filled with synchrotron-emitting cosmic ray electrons. Polarised emission from background sources can illuminate the foreground magneto-ionic media through the Faraday rotation effect, allowing the study of physical conditions in the intervening magnetised plasma.

Radio polarimetric observations of background extragalactic radio sources (EGSs) have been successful in revealing the magnetic fields in foreground astrophysical objects, such as discrete H~{\sc ii} regions in the Milky Way \citep{harvey-smith11,purcell15,costa16}, Galactic high velocity clouds \citep{mcclure-griffiths11,hill13,betti19}, the Galactic disk \citep{vaneck11}, the Galactic halo \citep{mao10,mao12,terral17}, the Magellanic system \citep{gaensler05,mao08,kaczmarek17}, nearby galaxies such as M31 \citep{han98,giessuebel13}, and cosmologically distant galaxies \citep{mao17}. As the polarised radiation traverses through the foreground media, its polarisation position angle (PA; [rad]) will be rotated by
\begin{equation}
\Delta{\rm PA} = \left[ 0.81 \int_\ell^0 n_e(s) B_\parallel(s)\,{\rm d}s \right] \cdot \lambda^2 \equiv \phi \cdot \lambda^2{\rm ,}
\end{equation}
where $\ell$ [pc] is the (physical) distance of the emitting volume from the observer, $n_e$ [${\rm cm}^{-3}$] is the electron density, $B_\parallel$ [$\mu$G] is the strength of the magnetic field component along the line of sight ($s$ [pc]; increasing away from the observer), $\lambda$ [m] is the wavelength of the electromagnetic wave, and $\phi$ [${\rm rad\,m}^{-2}$] is the Faraday depth (FD) of the emission region. This Faraday rotation effect encodes the physical conditions of the foreground magneto-ionic media, in particular $n_e$ and $B_\parallel$, into FD. The traditional way to extract the FD values of polarised sources is by PA measurements at two or more distinct frequency bands and perform a linear fit to PA against $\lambda^2$. In this case, FD is commonly referred to as Rotation Measure (RM) instead, which is the slope of the resulting fit. For situations where PA measurements are only available at two frequencies, the resulting FD (or RM) values can be ambiguous because wrapping(s) of PA can occur between the two bands. This is the so-called $n\pi$-ambiguity problem, and can be best mitigated by additional PA measurements at other frequency bands.

Modern radio telescopes equipped with broadband backends, such as the Karl G.\ Jansky Very Large Array (VLA), have started a new era in the study of cosmic magnetism. They opened up the possibility of spectro-polarimetric observations with unprecedented bandwidths (e.g.\ 1--2\,GHz in L-band and 2--4\,GHz in S-band for the VLA) and fine frequency resolutions (1--2\,MHz in the above-mentioned bands). This allows a simple eradication of $n\pi$-ambiguity in FD (or RM) measurements, since PAs at hundreds or even thousands of closely spaced frequencies can be measured simultaneously, ensuring no wrappings of PA between the channels. The even more important aspect of broadband spectro-polarimetric studies is the possibility to apply analysis methods such as RM-Synthesis \citep{brentjens05} and Stokes \textit{QU}-fitting \citep[e.g.][]{farnsworth11,osullivan12}. The former makes use of the Fourier-like behaviour of polarisation signal, such that input complex polarisation ($P = Q + iU$) as a function of $\lambda^2$ can be transformed into output Faraday spectrum ($F$; which is the complex polarisation as a function of $\phi$):
\begin{align}
P(\lambda^2) &= \int_{-\infty}^{+\infty} F(\phi) e^{2i\phi\lambda^2}\,{\rm d}\phi{\rm ,}\\
F(\phi) &= \int_{-\infty}^{+\infty} P(\lambda^2) e^{-2i\phi\lambda^2}\,{\rm d}\lambda^2{\rm .}
\end{align}
The latter technique is to fit the observed Stokes \textit{Q} and \textit{U} values as a function of $\lambda^2$ by using models of magnetised plasma along the line of sight. Both of the techniques allow exploration of Faraday complex sources \citep[e.g.][]{burn66,sokoloff98}, which emit at multiple FDs. These sources have varying polarisation fractions as a function of $\lambda^2$, and sometimes deviate from the linear relationship between PA and $\lambda^2$. Given sufficient $\lambda^2$ coverage, these sources would exhibit multiple peaks and/or extended component(s) in Faraday spectra. In contrast, Faraday simple sources emit at a single FD only, with constant polarisation fractions across $\lambda^2$, and have PA values varying linearly with $\lambda^2$. RM-Synthesis and \textit{QU}-fitting are widely used in broadband radio polarisation studies, with a growing success in revealing the Faraday complexities of a significant number of the observed EGSs \citep[e.g.][]{law11,anderson15,anderson16,kim16,osullivan17,kaczmarek18,pasetto18,schnitzeler19}.

\begin{table*}
\caption{Summary of the VLA Observations on 2014 July 03 \label{table:obs}}
\begin{tabular}{ccccccccc}
\hline
Start Time & End Time & Flux and Bandpass & Leakage & Phase & Target Source & Target Source & Angular \\
(UTC) & (UTC) & Calibrator & Calibrator & Calibrator & (NVSS) & (Other Name) & Resolution$^a$ \\
\hline
03:25:53 & 04:25:48 & 3C 286 & J1407+2827 & J1549+2125 & J154936+183500 & 4C +18.45 & $43^{\prime\prime} \times 40^{\prime\prime}$\\
&&&& J1623$-$1140 & J162706$-$091705 & --- & $59^{\prime\prime} \times 43^{\prime\prime}$\\
&&&&& J163927$-$124139 & --- & $64^{\prime\prime} \times 42^{\prime\prime}$\\
&&&& J1733$-$1304 & J170934$-$172853 & --- & $84^{\prime\prime} \times 41^{\prime\prime}$\\
&&&& J1924+3329 & J190255+315942 & 3C 395 & $52^{\prime\prime} \times 44^{\prime\prime}$\\
13:44:31 & 14:44:23 & 3C 138 & J0319+4130 & J0238+1636 & J022915+085125 & --- & $52^{\prime\prime} \times 37^{\prime\prime}$\\
&&&& J2202+4216 & J220205+394913 & --- & $56^{\prime\prime} \times 39^{\prime\prime}$\\
&&&&& J220927+415834 & --- & $57^{\prime\prime} \times 38^{\prime\prime}$\\
&&&&& J224412+405715 & --- & $53^{\prime\prime} \times 38^{\prime\prime}$\\
&&&&& J224549+394122 & 3C 452 & $51^{\prime\prime} \times 38^{\prime\prime}$\\
&&&& J2340+1333 & J234033+133300 & 4C +13.88 & $49^{\prime\prime} \times 41^{\prime\prime}$\\
&&&&& J235728+230226 & 4C +22.65 & $46^{\prime\prime} \times 38^{\prime\prime}$\\
22:28:10 & 23:57:59 & 3C 286 & J0713+4349 & J0837$-$1951 & J083930$-$240723 & --- & $102^{\prime\prime} \times 41^{\prime\prime}$\phantom{0}\\
&&&&& J084600$-$261054 & --- & $109^{\prime\prime} \times 40^{\prime\prime}$\phantom{0}\\
&&&&& J084701$-$233701 & --- & $101^{\prime\prime} \times 42^{\prime\prime}$\phantom{0}\\
&&&& J0921$-$2618 & J090015$-$281758 & --- & $116^{\prime\prime} \times 40^{\prime\prime}$\phantom{0}\\
&&&&& J091145$-$301305 & --- & $127^{\prime\prime} \times 39^{\prime\prime}$\phantom{0}\\
&&&&& J092410$-$290606 & --- & $114^{\prime\prime} \times 38^{\prime\prime}$\phantom{0}\\
&&&&& J093349$-$302700 & --- & $124^{\prime\prime} \times 38^{\prime\prime}$\phantom{0}\\
&&&& J1018$-$3144 & J093544$-$322845 & --- & $139^{\prime\prime} \times 38^{\prime\prime}$\phantom{0}\\
&&&&& J094750$-$371528 & --- & $177^{\prime\prime} \times 36^{\prime\prime}$\phantom{0}\\
&&&&& J094808$-$344010 & --- & $148^{\prime\prime} \times 36^{\prime\prime}$\phantom{0}\\
&&&& J1120+1420 & J111857+123442 & 4C +12.39 & $50^{\prime\prime} \times 40^{\prime\prime}$\\
\hline
\multicolumn{8}{l}{$^a$From channel maps at 1.5\,GHz}
\end{tabular}
\end{table*}

The largest RM catalogue of polarised radio sources to date is the \citet[][hereafter \citetalias{taylor09}]{taylor09} catalogue, which contains RM values of 37,543 radio sources north of $\delta = -40^\circ$ at a source density of higher than one per square degree. This makes it invaluable for the study of cosmic magnetism \citep[e.g.][]{stil11,oppermann12,purcell15,terral17}. \citetalias{taylor09} constructed the catalogue by re-analysing the NRAO VLA Sky Survey \citep[NVSS;][]{condon98} data, and thus it is also called the NVSS RM catalogue. While in the original NVSS catalogue the two intermediate frequencies (IFs; centred at 1364.9 and 1435.1\,MHz with bandwidths of 42\,MHz each) were combined, \citetalias{taylor09} processed data from the two IFs independently, allowing determination of RM from these two frequency bands. However, these RM values could then be susceptible to $n\pi$-ambiguity as discussed above. For each of the sources in their catalogue, the authors compared the observed amount of depolarisation with that expected from bandwidth depolarisation at the different allowed RM values, and also used the RM values of neighbouring sources within $3^\circ$, to minimise $n\pi$-ambiguity. However, it is not clear how effective this method really is at picking the correct RM values. Understanding the limits of the NVSS RM catalogue is vital to the study of cosmic magnetism. While upcoming polarisation surveys such as Polarization Sky Survey of the Universe's Magnetism \citep[POSSUM;][]{gaensler10} in 1130--1430\,MHz and VLA Sky Survey \citep[VLASS;][]{myers14} in 2--4\,GHz are expected to bring vastly higher RM densities compared to \citetalias{taylor09}, the two surveys either do not have exact sky or frequency coverage as \citetalias{taylor09}. This means the NVSS RM catalogue will remain a unique dataset for studying the magnetised Universe, complementing the VLASS in the frequency domain and POSSUM in the sky domain, in addition to both in the time domain. A prior deeper understanding in the systematics of \citetalias{taylor09} will facilitate future robust comparisons among these surveys. The focus of our work here is to effectively test the reliability of the \citetalias{taylor09} RM values by validating a small sample of \citetalias{taylor09} sources using broadband polarimetry, which provides us with $n\pi$-ambiguity-free FD.

In this paper, we report the results from new broadband observations of 23 candidates from the NVSS RM catalogue which could suffer from $n\pi$-ambiguity. The observational setup and data reduction procedures are described in Section~\ref{sec:obs}, and the results are presented in Section~\ref{sec:results}. In Section~\ref{sec:discussion}, we discuss the implications of the results on the $n\pi$-ambiguity in the \citetalias{taylor09} catalogue, and also explore the Faraday complexities of the targets revealed by the new broadband observations. Finally, we conclude this work in Section~\ref{sec:conclusion}. In the companion Paper II \citep{ma19b}, we further compare this dataset with the \citetalias{taylor09} catalogue in matching frequency ranges to quantify the effects of the off-axis instrumental polarisation on the \citetalias{taylor09} RM measurements. Throughout the paper, we adopt a cosmology in accordance to the latest \textit{Planck} results \citep[i.e., $H_0 = 67.8\,{\rm km\,s}^{-1}\,{\rm Mpc}^{-1}$ and $\Omega_m = 0.308$;][]{planck}. 

\newpage
\section{OBSERVATIONS AND DATA REDUCTION} \label{sec:obs}
\subsection{New Observations and Calibration}
We selected the 23 target sources from the \citetalias{taylor09} catalogue. They have high $|{\rm RM}_{\rm TSS09}| \gtrsim 300\,{\rm rad\,m}^{-2}$ and are situated away from the Galactic plane ($|b| > 10^\circ$)\footnote{Except for J234033$+$133300, which has ${\rm RM}_{\rm TSS09} = +56.7 \pm 6.3\,{\rm rad\,m}^{-2}$. This source was also observed because it was thought to have a high emission measure (EM; $\sim 140\,{\rm cm}^{-6}\,{\rm pc}$) but low $|{\rm RM}|$, which could be another manifestation of $n\pi$-ambiguity. However, upon close examination after the observation was conducted, we found that the EM along this sightline is actually $\lesssim 10\,{\rm cm}^{-6}\,{\rm pc}$, thus disqualifying this source as an $n\pi$-ambiguity candidate. This source will not be included in the statistical analysis on $n\pi$-ambiguity in this work. However, we later found that this source is unpolarised, which leads to implications on the residual off-axis polarisation leakage of \citetalias{taylor09} (see Paper II).}. In this region, the Galactic FD (or RM) contributions are less significant, with $\approx 99$ per cent of the \citetalias{taylor09} sources with $|{\rm RM}_{\rm TSS09}| < 150\,{\rm rad\,m}^{-2}$. The peculiar population we selected, with high $|{\rm RM}_{\rm TSS09}|$, could be statistical \textit{outliers} from the generally low $|{\rm RM}_{\rm TSS09}|$ population, either because they have high intrinsic FD (or RM) values or they are positioned along special lines of sight with high foreground FD (or RM) contributions. On the other hand, our target sources could also be \textit{out-liars} with erroneous ${\rm RM}_{\rm TSS09}$ values, deviating from the true RM by multiples of $\pm 652.9\,{\rm rad\,m}^{-2}$ due to $n\pi$-ambiguity \citepalias{taylor09} and causing them to stand out from the majority. However, we note that our selection criteria does not allow us to study sources with high true $|{\rm RM}|$ having low reported $|{\rm RM}_{\rm TSS09}|$ due to $n\pi$-ambiguity, and thus our study here only focuses on cases where sources with low true $|{\rm RM}|$ are ``boosted'' to high $|{\rm RM}_{\rm TSS09}|$ due to $n\pi$-ambiguity. We further selected only bright sources with NVSS total intensities larger than 100\,mJy to ensure that sufficient signal-to-noise ratio could be achieved.

Our new broadband data were acquired using the VLA in L-band ($1$--$2\,{\rm GHz}$) in D array configuration. The observations were carried out on 2014 July 03 in three observing blocks, and are summarised in Table~\ref{table:obs} where the observing time, calibrators, target sources, and angular resolutions are listed. For each of the target sources, the integration time is about $3$--$4\,{\rm minutes}$. We used the Common Astronomy Software Applications (CASA) package \citep[version 4.4.0; ][]{mcmullin07} for all of the data reduction procedures. 

The three measurement sets were calibrated independently. Hanning smoothing is first applied to all the visibilities in frequency domain to remove the Gibbs phenomenon, and the antenna position calibration is applied to the dataset. Then, we flagged out times when the antennas were not performing as intended or when prominent radio frequency interferences (RFI) were seen. Next, we determined the delay, bandpass, and gain solutions using the flux and/or phase calibrators, with the absolute flux densities following the \cite{perley13a} scales. The PA calibration was done by using the previously determined PAs of the flux calibrators 3C 286 and 3C 138 \citep{perley13b}, while the on-axis instrumental leakage was corrected for by observing standard unpolarised leakage calibrators (see Table~\ref{table:obs}). Finally, we applied one round of phase self calibration to all our target sources to further improve the gain solution.

\begin{table}
\caption{Positions of Individual Components of the Spatial Doubles \label{table:double}}
\begin{tabular}{ccc}
\hline
Source & Right Ascension & Declination \\
(NVSS) & (J2000; h m s) & (J2000; $^\circ$ $^\prime$ $^{\prime\prime}$) \\
\hline
J091145$-$301305 && \\
\multicolumn{1}{c}{$\cdots$ a} & 09 11 42.47 $\pm$ 0.04 & $-$30 13 19.26 $\pm$ 1.45 \\
\multicolumn{1}{c}{$\cdots$ b} & 09 11 46.33 $\pm$ 0.02 & $-$30 12 58.63 $\pm$ 0.74 \\
J092410$-$290606 && \\
\multicolumn{1}{c}{$\cdots$ a} & 09 24 10.09 $\pm$ 0.02 & $-$29 05 45.36 $\pm$ 0.79 \\
\multicolumn{1}{c}{$\cdots$ b} & 09 24 11.44 $\pm$ 0.02 & $-$29 06 26.66 $\pm$ 0.75 \\
J093544$-$322845 && \\
\multicolumn{1}{c}{$\cdots$ a} & 09 35 43.98 $\pm$ 0.02 & $-$32 28 48.51 $\pm$ 0.65 \\
\multicolumn{1}{c}{$\cdots$ b} & 09 35 43.79 $\pm$ 0.02 & $-$32 29 40.03 $\pm$ 0.60 \\
J162706$-$091705 && \\ 
\multicolumn{1}{c}{$\cdots$ a} & 16 27 04.53 $\pm$ 0.02 & $-$09 16 55.99 $\pm$ 0.64 \\
\multicolumn{1}{c}{$\cdots$ b} & 16 27 06.78 $\pm$ 0.01 & $-$09 17 06.50 $\pm$ 0.20 \\
J163927$-$124139 && \\
\multicolumn{1}{c}{$\cdots$ a} & 16 39 27.09 $\pm$ 0.01 & $-$12 41 26.41 $\pm$ 0.15 \\
\multicolumn{1}{c}{$\cdots$ b} & 16 39 28.20 $\pm$ 0.01 & $-$12 42 09.07 $\pm$ 0.27 \\
\hline
\end{tabular}
\end{table}

\subsection{Full L-Band Images} \label{sec:fullbandreduction}
With the calibrated visibilities, we formed a series of Stokes \textit{I}, \textit{Q}, and \textit{U} images for each target source at different frequencies across L-band, combining $4\,{\rm MHz}$ of visibility data to form the images for each step in the frequency axis. The Clark deconvolution algorithm in CASA task \texttt{CLEAN} was adopted, with Briggs visibilities weighting of \texttt{robust} $= 0$ \citep{briggs95}. We did not further smooth the resulting images, as it would only be necessary if we directly combine images at different frequencies. We list the angular resolution at 1.5\,GHz of each pointing in Table~\ref{table:obs}. At the spatial resolution of our observations, our targets can be divided into three morphology classes -- single (unresolved), double (resolved into two unresolved components), and extended. The typical root-mean-square (rms) noise of each $4\,{\rm MHz}$ image is about $1.6\,{\rm mJy\,beam}^{-1}$ in Stokes \textit{I}, and $1.0\,{\rm mJy\,beam}^{-1}$ in Stokes \textit{Q} and \textit{U}. 

We measured the Stokes \textit{I}, \textit{Q}, and \textit{U} values of our target sources per frequency step. We used different methods depending on whether the sources are spatially resolved with our observational setup. For spatial singles and doubles, we used the CASA task \texttt{IMFIT} to extract the flux densities and their uncertainties. The full-width at half-maximum (FWHM) of the Gaussian components are fixed as that of the synthesised beam at each frequency step, and the fitted source locations in Stokes \textit{I} are also used for Stokes \textit{Q} and \textit{U}. The positions of the individual components of the five double sources in our sample are listed in Table~\ref{table:double}. For extended sources (J094750$-$371528 and J224549$+$394122), we used the multi-frequency synthesis (MFS) algorithm with \texttt{nterms}~$=2$ (which incorporates the spectral indices of the sources) to form Stokes \textit{I} images using the entire L-band for each of the sources, from which $6\sigma$ contours in Stokes \textit{I} enclosing the target sources are defined. The CASA task \texttt{IMSTAT} is then used to integrate the Stokes \textit{I}, \textit{Q}, and \textit{U} flux densities within the contour for each channel map. We note that using integrated flux densities discards all the spatial information we have of these two sources, and may increase Faraday complexity and/or cause beam depolarisation. A detailed spatial analysis of them is included in Appendix~\ref{sec:spatial}. The radio spectra of our targets are reported in Paper II, in which we address the potential Stokes \textit{I} and RM time variabilities of our sample.

\begin{table*}
\caption{Results of RM-Synthesis on Broadband VLA Data \label{table:rmsyn}}
\begin{tabular}{lcccccccc}
\hline
\multicolumn{1}{c}{Source} & $p$ & ${\rm PA}_0$ & $\phi$ & $\overline\phi$ & RM$_{\rm TSS09}$ & $|\overline\phi - {\rm RM}_{\rm TSS09}|$ & $\delta\phi$ & $\delta\phi_0$ \\
\multicolumn{1}{c}{(NVSS)} & (\%) & ($^\circ$) & (${\rm rad\,m}^{-2}$) & (${\rm rad\,m}^{-2}$) & (${\rm rad\,m}^{-2}$) & (${\rm rad\,m}^{-2}$) & (${\rm rad\,m}^{-2}$) & (${\rm rad\,m}^{-2}$) \\
\hline
\multicolumn{9}{c}{\textbf{Outliers (Reliable $\mathbf{RM}_\mathbf{TSS09}$)}}\\
\hline
J083930$-$240723 & $4.54^{+0.06}_{-0.05}$  & $+5.8^{+2.2}_{-2.2}$ & $+325.9\pm 1.0$ & $+325.9\pm 1.0$ & \phantom{0}$+345.2\pm 10.5$ & \phantom{0}19.3 & $128.0$ & $128.0$ \\
J084701$-$233701 & $3.12^{+0.10}_{-0.10}$  & $-88.5^{+4.4}_{-4.4}$ & $+384.8\pm 2.0$ & $+384.8\pm 2.0$ & \phantom{0}$+429.5\pm 15.3$ & \phantom{0}44.7 & $128.0$ & $128.0$ \\
J090015$-$281758 & $4.44^{+0.02}_{-0.02}$  & $+6.8^{+0.5}_{-0.5}$ & $+352.1\pm 0.2$ & $+352.1\pm 0.2$ & $+320.6\pm 4.2$ & \phantom{0}31.5 & \phantom{0}$76.0$ & \phantom{0}$76.0$ \\
J092410$-$290606$^{\star\star}$ & --- & --- & --- & $+527.6^{+0.3}_{-0.3}$ & $+472.9\pm 6.2$ & \phantom{0}54.7 & --- & \phantom{0}76.0 \\
\multicolumn{1}{c}{$\cdots$ a} & $8.23^{+0.06}_{-0.06}$ & $+60.1^{+0.6}_{-0.6}$ & $+526.1\pm 0.3$ & --- & --- & --- & \phantom{0}$76.0$ & --- \\
\multicolumn{1}{c}{$\cdots$ b} & $3.80^{+0.09}_{-0.09}$ & $+74.8^{+1.9}_{-1.9}$ & $+530.8\pm 0.8$ & --- & --- & --- & \phantom{0}$76.0$ & --- \\
J093349$-$302700 & $6.03^{+0.08}_{-0.08}$  & $-33.6^{+1.7}_{-1.7}$ & $+341.6\pm 0.8$ & $+341.6\pm 0.8$ & $+313.4\pm 7.7$ & \phantom{0}28.2 & $112.0$ & $112.0$ \\
J093544$-$322845$^{\star\star}$ & --- & --- & --- & $+390.9^{+0.3}_{-0.3}$ & $+368.1\pm 9.3$ & \phantom{0}22.8 & --- & \phantom{0}76.0 \\
\multicolumn{1}{c}{$\cdots$ a} & $7.68^{+0.07}_{-0.07}$ & $-50.2^{+0.7}_{-0.7}$ & $+393.8\pm 0.3$ & --- & --- & --- & \phantom{0}$76.0$ & --- \\
\multicolumn{1}{c}{$\cdots$ b} & $5.16^{+0.08}_{-0.08}$ & $+74.2^{+1.3}_{-1.3}$ & $+387.2\pm 0.6$ & --- & --- & --- & \phantom{0}$76.0$ & --- \\
J094750$-$371528$^{\odot\ddagger}$ & $3.90^{+0.25}_{-0.24}$  & $-8.7^{+5.2}_{-5.2}$ & $+328.8\pm 2.2$ & $+328.8\pm 2.2$ & $+311.0\pm 7.2$ & \phantom{0}17.8 & $113.6\pm 0.4$ & \phantom{0}$77.0$ \\
J162706$-$091705$^{\star\star}$ & --- & --- & --- & $-327.8 \pm 0.7$ & \phantom{0}$-297.2\pm 12.8$ & \phantom{0}30.6 & --- & 104.0 \\
\multicolumn{1}{c}{$\cdots$ a} & --- & --- & --- & --- & --- & --- & --- & --- \\
\multicolumn{1}{c}{$\cdots$ b} & $11.32^{+0.16}_{-0.15}$ & $+4.4^{+1.4}_{-1.4}$ & $-327.8\pm 0.7$ & --- & --- & --- & $104.0$ & --- \\
J163927$-$124139$^{\star\star}$ & --- & --- & --- & $-331.4^{+0.3}_{-0.3}$ & $-313.5\pm 3.6$ & \phantom{0}17.9 & --- & 104.0 \\
\multicolumn{1}{c}{$\cdots$ a} & $9.11^{+0.06}_{-0.06}$ & $+32.1^{+0.8}_{-0.8}$ & $-328.4\pm 0.3$ & --- & --- & --- & $104.0$ & --- \\
\multicolumn{1}{c}{$\cdots$ b} & $10.10^{+0.07}_{-0.07}$ & $+68.6^{+0.8}_{-0.8}$ & $-336.0\pm 0.4$ & --- & --- & --- & $104.0$ & --- \\
J220205$+$394913 & $8.39^{+0.10}_{-0.10}$  & $+59.1^{+1.0}_{-1.0}$ & $-367.2\pm 0.4$ & $-367.2\pm 0.4$ & $-349.1\pm 6.6$ & \phantom{0}18.1 & \phantom{0}$76.0$ & \phantom{0}$76.0$ \\
J220927$+$415834 & $6.92^{+0.04}_{-0.04}$  & $-12.5^{+0.5}_{-0.5}$ & $-338.1\pm 0.2$ & $-338.1\pm 0.2$ & $-336.0\pm 5.4$ & \phantom{00}2.1 & \phantom{0}$76.0$ & \phantom{0}$76.0$ \\
\hline
\multicolumn{9}{c}{\textbf{Out-\textit{liars} ($n\pi$-ambiguity)}}\\
\hline
J022915$+$085125$^{!\ddagger}$ & --- & --- & --- & \phantom{0}$+13.6 \pm 1.0$ & $+521.2\pm 8.0$ & 507.6 & --- & $124.0$ \\
\multicolumn{1}{c}{$\cdots$ FC 1} & $0.27^{+0.03}_{-0.03}$  & $+63.2^{+12.9}_{-13.1}$ & $-246.8\pm 5.9$  & --- & --- & --- & \phantom{0}$74.7\pm 4.8$ & --- \\
\multicolumn{1}{c}{$\cdots$ FC 2} & $4.05^{+0.07}_{-0.07}$  & $-24.2^{+2.2}_{-2.2}$ & \phantom{0}$+13.6\pm 1.0$  & --- & --- & --- & $187.3\pm 0.2$ & --- \\
\multicolumn{1}{c}{$\cdots$ FC 3} & $0.27^{+0.03}_{-0.03}$  & $+69.7^{+12.9}_{-12.8}$ & $+272.8\pm 5.8$  & --- & --- & --- & \phantom{0}$72.6\pm 4.4$ & --- \\
J091145$-$301305$^{\star\star}$ & --- & --- & --- & $+246.9^{+0.3}_{-0.3}$ & $-426.1\pm 3.5$ & 673.0 & --- & \phantom{0}76.0 \\
\multicolumn{1}{c}{$\cdots$ a} & $8.68^{+0.21}_{-0.20}$ & $-27.5^{+2.1}_{-2.1}$ & $+245.0\pm 0.9$ & --- & --- & --- & \phantom{0}$76.0$ & --- \\
\multicolumn{1}{c}{$\cdots$ b} & $17.15^{+0.11}_{-0.11}$ & $-15.3^{+0.6}_{-0.6}$ & $+247.4\pm 0.3$ & --- & --- & --- & \phantom{0}$76.0$ & --- \\
J094808$-$344010$^\ddagger$ & --- & --- & --- & $+382.7^{+2.6}_{-2.4}$ & \phantom{0}$-327.9\pm 10.6$ & 710.6 & --- & \phantom{0}$76.0$ \\
\multicolumn{1}{c}{$\cdots$ FC 1} & $4.20^{+0.08}_{-0.08}$  & $+53.8^{+2.0}_{-2.0}$ & $+364.7\pm 0.9$  & --- & --- & --- & \phantom{0}$76.0$ & --- \\
\multicolumn{1}{c}{$\cdots$ FC 2} & $0.73^{+0.11}_{-0.10}$  & $-71.6^{+10.7}_{-11.0}$ & $+486.0\pm 5.2$  & --- & --- & --- & \phantom{0}$76.0$ & --- \\
J111857$+$123442$^\ddagger$ & --- & --- & --- & $+79.4^{+2.6}_{-2.8}$ & $-465.4\pm 5.7$ & 544.8 & --- & \phantom{0}$76.0$ \\
\multicolumn{1}{c}{$\cdots$ FC 1} & $0.15^{+0.02}_{-0.02}$  & $-59.8^{+10.3}_{-10.8}$ & \phantom{00}$-2.5\pm 5.2$  & --- & --- & --- & \phantom{0}$76.0$ & --- \\
\multicolumn{1}{c}{$\cdots$ FC 2} & $0.63^{+0.02}_{-0.02}$  & $+51.7^{+2.5}_{-2.5}$ & \phantom{0}$+98.8\pm 1.2$  & --- & --- & --- & \phantom{0}$76.0$ & --- \\
J170934$-$172853$^\ddagger$ & --- & --- & --- & $+106.2^{+1.8}_{-1.9}$ & \phantom{0}$-490.0\pm 12.7$ & 596.2 & --- & \phantom{0}$76.0$ \\
\multicolumn{1}{c}{$\cdots$ FC 1} & $0.43^{+0.06}_{-0.05}$  & $-3.6^{+10.1}_{-10.2}$ & \phantom{0}$-26.5\pm 4.8$  & --- & --- & --- & \phantom{0}$76.0$ & --- \\
\multicolumn{1}{c}{$\cdots$ FC 2} & $3.86^{+0.05}_{-0.05}$  & $-37.9^{+1.1}_{-1.1}$ & $+120.9\pm 0.5$  & --- & --- & --- & \phantom{0}$76.0$ & --- \\
J190255$+$315942$^\ddagger$ & --- & --- & --- & $+142.2^{+1.1}_{-1.1}$ & $-424.3\pm 2.6$ & 566.5 & --- & \phantom{0}$76.0$ \\
\multicolumn{1}{c}{$\cdots$ FC 1} & $1.34^{+0.05}_{-0.05}$  & $-25.2^{+3.0}_{-3.0}$ & \phantom{0}$+84.9\pm 1.4$  & --- & --- & --- & \phantom{0}$76.0$ & --- \\
\multicolumn{1}{c}{$\cdots$ FC 2} & $2.45^{+0.05}_{-0.05}$  & $+17.4^{+1.6}_{-1.6}$ & $+173.4\pm 0.8$  & --- & --- & --- & \phantom{0}$76.0$ & --- \\
J224412$+$405715 & $3.55^{+0.05}_{-0.05}$  & $+84.5^{+1.3}_{-1.3}$ & $-320.4\pm 0.6$ & $-320.4\pm 0.6$ & \phantom{0}$+345.8\pm 14.4$ & 666.2 & \phantom{0}$76.0$ & \phantom{0}$76.0$ \\
J224549$+$394122$^{\odot\ddagger}$ & --- & --- & --- & $-278.6^{+0.8}_{-1.0}$ & $+373.3\pm 6.4$ & 651.9 & --- & \phantom{0}$76.0$ \\
\multicolumn{1}{c}{$\cdots$ FC 1} & $0.20^{+0.04}_{-0.03}$  & $+72.3^{+15.9}_{-15.9}$ & $-407.3\pm 6.3$  & --- & --- & --- & $110.9\pm 9.6$ & --- \\
\multicolumn{1}{c}{$\cdots$ FC 2} & $5.37^{+0.02}_{-0.02}$  & $+4.5^{+0.4}_{-0.4}$ & $-273.8\pm 0.2$  & --- & --- & --- & \phantom{0}$80.5\pm 0.0$ & --- \\
J235728$+$230226 & $0.20^{+0.02}_{-0.02}$  & $+46.5^{+13.6}_{-13.6}$ & \phantom{0}$+42.3\pm 6.1$ & \phantom{0}$+42.3\pm 6.1$ & \phantom{0}$-556.7\pm 13.8$ & 599.0 & $124.0$ & $124.0$ \\
\hline
\multicolumn{9}{c}{\textbf{Others}}\\
\hline
J084600$-$261054$^\times$ & --- & --- & --- &--- & \phantom{0}$+481.0\pm 11.4$ & --- & --- & $128.0$ \\
J154936$+$183500$^{?\ddagger}$ & --- & --- & --- & $-119.3^{+7.5}_{-7.4}$ & \phantom{0}$-426.8\pm 14.6$ & 307.5 & --- & \phantom{0}$76.0$ \\
\multicolumn{1}{c}{$\cdots$ FC 1} & $0.50^{+0.03}_{-0.03}$  & $+23.2^{+4.9}_{-4.9}$ & $-315.3\pm 2.1$  & --- & --- & --- & \phantom{0}$87.9\pm 0.3$ & --- \\
\multicolumn{1}{c}{$\cdots$ FC 2} & $0.31^{+0.03}_{-0.03}$  & $+77.1^{+8.3}_{-8.3}$ & \phantom{0}$-31.8\pm 3.5$  & --- & --- & --- & \phantom{0}$90.8\pm 3.1$ & --- \\
\multicolumn{1}{c}{$\cdots$ FC 3} & $0.35^{+0.03}_{-0.03}$  & $-48.2^{+6.6}_{-6.6}$ & \phantom{0}$+81.3\pm 2.8$  & --- & --- & --- & \phantom{0}$84.4\pm 1.7$ & --- \\
J234033$+$133300$^\times$ & --- & --- & --- &--- & \phantom{0}$+56.7\pm 6.3$ & --- & --- & $124.0$ \\
\hline
\multicolumn{9}{l}{$^\times$Unpolarised sources} \\
\multicolumn{9}{l}{$^?$Special case compared to \citetalias{taylor09} catalogue (see Section~\ref{sec:j1549})} \\
\multicolumn{9}{l}{$^{\star\star}$Double point sources} \\
\multicolumn{9}{l}{$^\odot$Extended sources} \\
\multicolumn{9}{l}{$^!$Faraday components (FCs) 1 and 3 may be artefacts corresponding to RMTF sidelobes (see text)} \\
\multicolumn{9}{l}{$^\ddagger$Faraday complex from RM-Synthesis} \\
\end{tabular}
\end{table*}

\begin{figure*}
\includegraphics[width=\doublecolumnwidth]{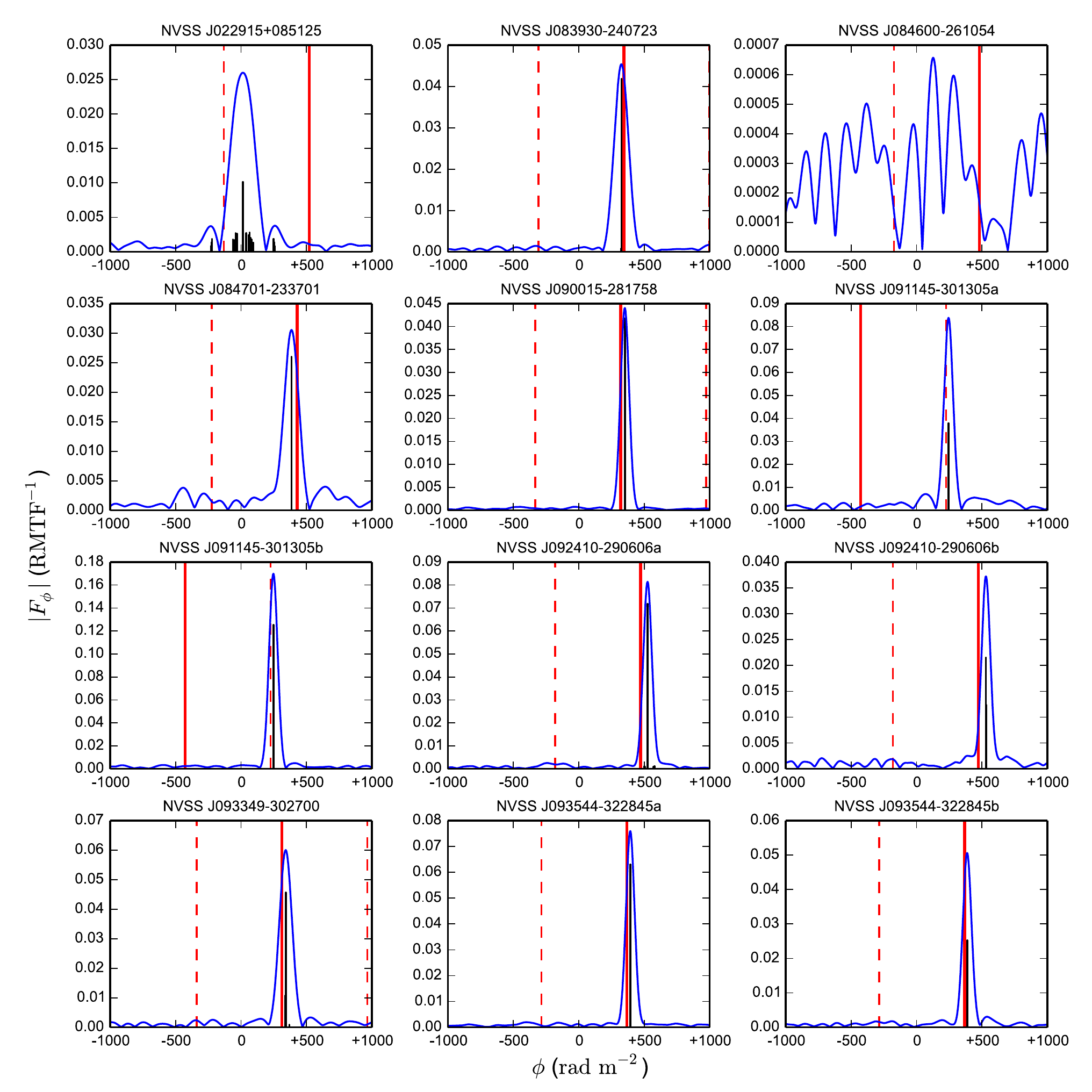}
\caption{Faraday spectra of our target sources. Blue lines show the amplitude of the complex Faraday spectra after deconvolution, with the black bars representing the clean components. The RM values from the NVSS RM catalogue \citepalias{taylor09} are represented by the red vertical solid lines, while those RM values corresponding to $\pm 1\pi$-ambiguity are indicated by red vertical dashed lines. We only show the spectra within the FD range of $-1000$ to $+1000\,{\rm rad\,m}^{-2}$, as significant Faraday components are not found outside of this range. \label{fig:rmsyn}}
\end{figure*}

\begin{figure*}
\ContinuedFloat
\includegraphics[width=\doublecolumnwidth]{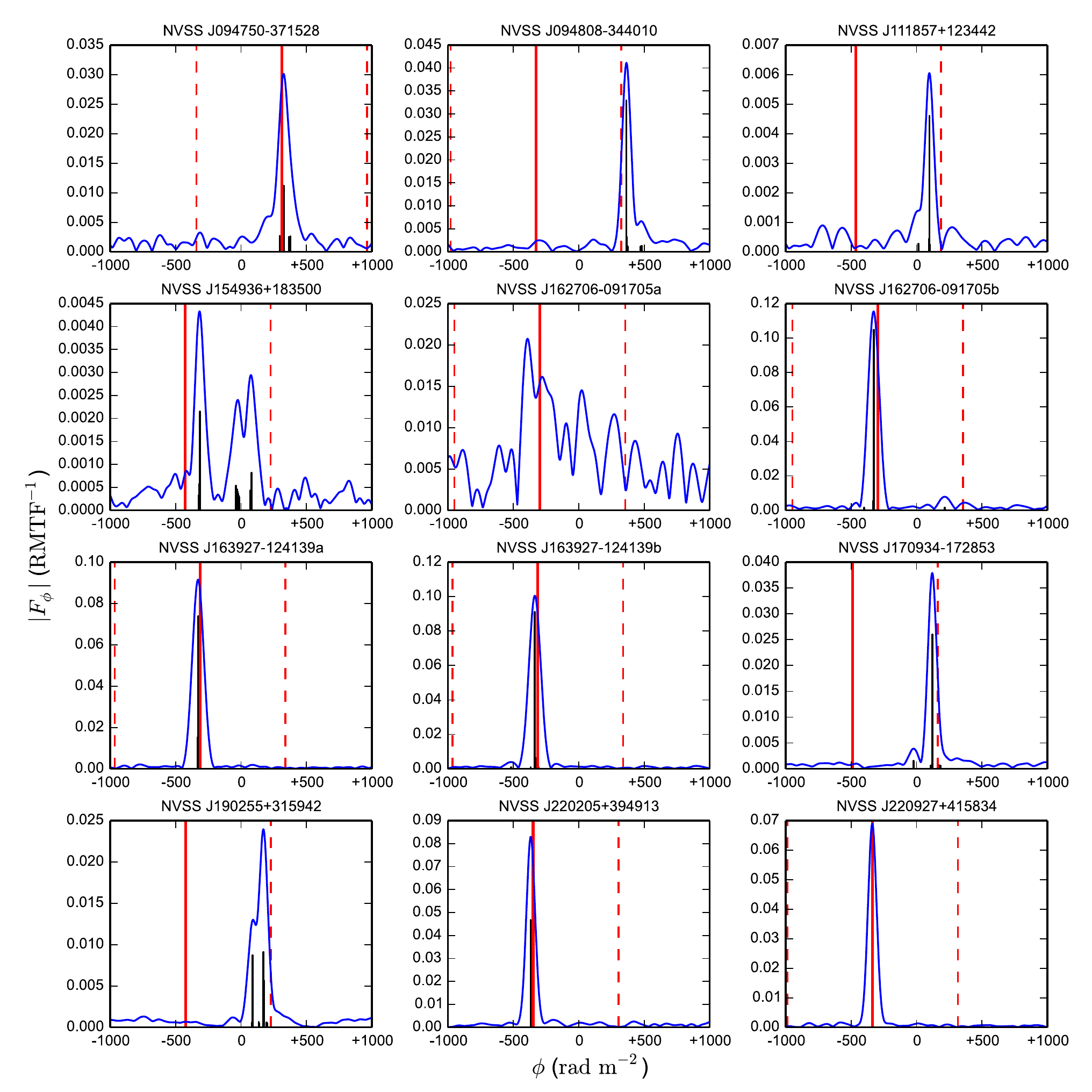}
\caption{(Continued) Faraday spectra of our target sources.}
\end{figure*}

\begin{figure*}
\ContinuedFloat
\includegraphics[width=\doublecolumnwidth]{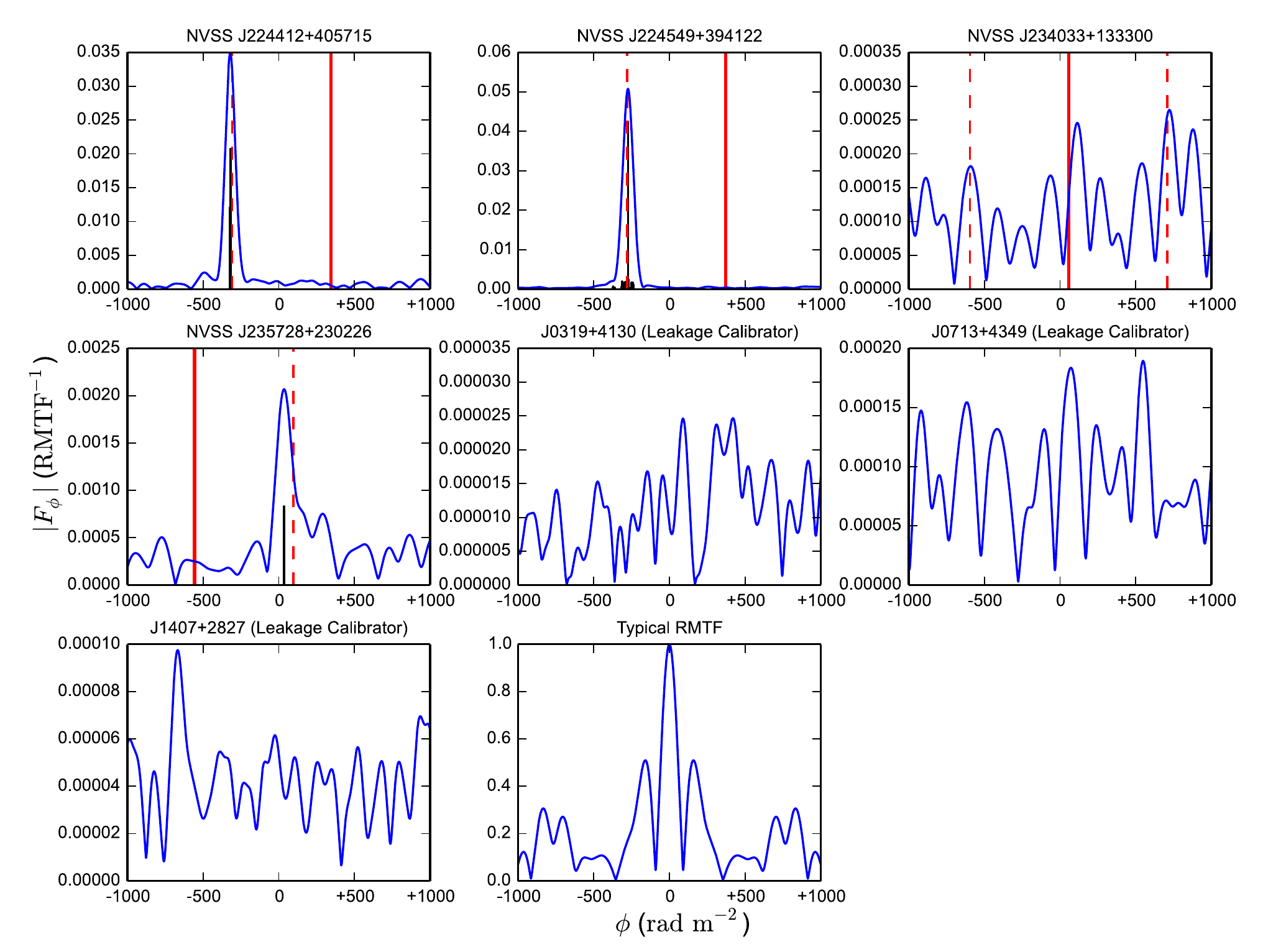}
\caption{(Continued) Faraday spectra of our target sources, as well as that of the leakage calibrators. The typical RMTF of our L-band observations is shown in the last panel.}
\end{figure*}

\section{BROADBAND SPECTRO-POLARIMETRIC ANALYSIS}\label{sec:results}
\subsection{Rotation Measure Synthesis \label{sec:rmsyn}}
Using the extracted Stokes \textit{I}, \textit{Q}, and \textit{U} values for every $4\,{\rm MHz}$ channel map (Section~\ref{sec:fullbandreduction}), we performed RM-Synthesis \citep{brentjens05} on all our target sources. For double sources, each of the spatial components are analysed independently. We used the Python-based RM-Synthesis code, \texttt{pyrmsynth}\footnote{Available on \url{http://www.github.com/mrbell/pyrmsynth}.}, to perform this analysis, including RM-Clean algorithm \citep[e.g.][]{heald09} to deconvolve the Faraday spectra. The $q=Q/I$ and $u=U/I$ values are used as the inputs, and therefore the resulting complex Faraday spectra (sometimes referred to as Faraday dispersion functions in the literature) are in units of polarisation fraction ($p$) per Rotation Measure Transfer Function (RMTF). With our observational setup, the resolution of Faraday spectrum, maximum detectable scale, and maximum detectable FD are \citep[equations~61--63 in][]{brentjens05}
\begin{gather}
\delta\phi_0 \approx \frac{2\sqrt{3}}{\Delta\lambda^2} \approx 76\text{--}128\,{\rm rad\,m}^{-2}{\rm ,} \\
\text{max-scale} \approx \frac{\pi}{\lambda_{\rm min}^2} \approx 143\,{\rm rad\,m}^{-2}\text{, and} \\
||\phi_{\rm max}|| \approx \frac{\sqrt{3}}{\delta \lambda^2} \approx (4\text{--}20) \times 10^3\,{\rm rad\,m}^{-2}{\rm ,}
\end{gather}
respectively. The quoted range for $\delta \phi_0$ is due to the slightly different $\lambda^2$ coverage for each source as the result of flagging (individual values listed in Table~\ref{table:rmsyn}), and that for $|| \phi_{\rm max} ||$ is because of the difference in widths of the $4\,{\rm MHz}$ channels in $\lambda^2$ space across L-band. We adopted a normalised inverse noise variance weighting function of \citep[e.g.][]{schnitzeler17}
\begin{equation}
W(\lambda^2) \propto \frac{1}{\sigma_{q}^2(\lambda^2) + \sigma_{u}^2(\lambda^2)}{\rm ,}
\end{equation}
where $\sigma_{q}$ and $\sigma_{u}$ are the uncertainties in $q$ and $u$ respectively. The Faraday spectra were formed within $-2000 \leq \phi$ (${\rm rad\,m}^{-2}$) $\leq +2000$ with a step size of $2\,{\rm rad\,m}^{-2}$. We first perform trial cleans to determine the rms noise (denoted as $\sigma$ here) in the source-free FD ranges of $|\phi| \geq 1000\,{\rm rad\,m}^{-2}$ from the $q_\phi$ and $u_\phi$ Faraday spectra. The final Faraday spectra are cleaned down to $6\sigma$ only so as to avoid over-cleaning, which can introduce artefacts to the resulting spectra. 

The Faraday spectra amplitudes ($|F_\phi| = \sqrt{q_\phi^2 + u_\phi^2}$) are shown in Figure~\ref{fig:rmsyn}. For each amplitude spectrum, we counted the number of peaks higher than $6\sigma$, and then we fitted the spectrum with the corresponding number of Gaussian components plus a $y$-offset to extract the FD values and widths of the peaks. This $6\sigma$ cutoff grants us an insignificant false detection rate of $\lesssim 0.5$ per cent \citep[e.g.][]{george12}, and a negligible Ricean polarisation bias \citep[$\lesssim 1.5$ per cent;][]{wardle74}. If the fitted FWHM ($\delta\phi$) of a peak is within 10 per cent from the theoretical RMTF FWHM value (i.e.\ $\delta\phi_0$; obtained from \texttt{pyrmsynth} output), we re-fit the spectrum with $\delta\phi$ being fixed at $\delta\phi_0$. The uncertainties in FD are obtained by \citep[e.g.][]{mao10,iacobelli13}
\begin{equation}
\frac{\delta\phi}{2\cdot(S/N)}{\rm ,}
\end{equation}
where $S/N$ is the signal-to-noise ratio of the peak. The FD and $\delta\phi$ of the peaks are then fixed and used to fit the $q_\phi$ and $u_\phi$ Faraday spectra to extract the complex polarisation of the Faraday components that they correspond to. The obtained values, namely $\phi$, $\delta\phi$, and complex polarisation, are then used to calculate $p$ and intrinsic PA (${\rm PA}_0$) of each Faraday component. The uncertainties are propagated by Monte Carlo simulations with $10^6$ realisations per source, starting from assuming that $q$, $u$, $\phi$, and $\delta\phi$ obtained from RM-Synthesis above follow Gaussian statistics. We evaluated the 68.3 per cent confidence interval (corresponding to $1\sigma$ under normal distribution), which are listed as the asymmetric errors in Table~\ref{table:rmsyn}. Such an error propagation method is needed, since strictly speaking the uncertainties of both $p$ and ${\rm PA}_0$ do not follow Gaussian distributions. A caveat to the results here is that the polarisation fraction $p$ is the polarised intensity of the Faraday component divided by the total intensity of the entire spatial component. The Ricean polarisation bias is not corrected for because it is insignificant at our signal-to-noise levels (see above).

We also formed Faraday spectra for the leakage calibrators J0319+4130, J0713+4349, and J1407+2827, in order to constrain the remaining instrumental polarisation leakage of our observations. Their spectra are also shown in Figure~\ref{fig:rmsyn}, with peak values of $0.003 \pm 0.001$, $0.019 \pm 0.006$, and $0.010 \pm 0.003$ per cent, respectively. Note that these values could be due to random noise fluctuations leading to polarisation bias \citep[e.g.][]{george12} instead of due to residual instrumental polarisation leakage, and are therefore upper limits to the actual remaining leakage levels of our calibrated data. We conclude that the residual polarisation leakage in our data is at $< 0.02$ per cent level.

One point to note is that for one of our sources, J022915+085125, Faraday components (FCs) 1 and 3 are likely artefacts corresponding to the sidelobes of the RMTF (see Table~\ref{table:rmsyn}), most likely because the main (physical) peak is Faraday thick, leading to sub-optimal deconvolution with RM-Clean. The two components are symmetric about the prominent Faraday component 2, having the same polarisation fraction of 0.27 per cent, ${\rm PA}_0$ and $\phi$ offsets from component 2 by about $90^\circ$ and $260\,{\rm rad\,m}^{-2}$ respectively, and $\delta\phi \approx 73\,{\rm rad\,m}^{-2}$, less than the theoretical value of $124\,{\rm rad\,m}^{-2}$. Upon inspection of the (complex) RMTF of this source, we find that the secondary maxima are offset from the primary by $161\,{\rm rad\,m}^{-2}$, phase offset by $180^\circ$ (i.e.\ $90^\circ$ in PA), and have FWHM of about $108\,{\rm rad\,m}^{-2}$. We have therefore ignored these two components in the remainder of this paper.

For the six sources that can be decomposed into multiple spatial or Faraday components, it is not trivial to directly compare the multiple FD values against the single ${\rm RM}_{\rm TSS09}$ value of each source. Therefore, we define a polarisation-weighted FD as
\begin{equation}
\overline\phi = \sum_i \frac{p_i \cdot S_{1.4\,{\rm GHz},i} \cdot \phi_i}{\sum_j p_j \cdot S_{1.4\,{\rm GHz},j}}{\rm ,}
\end{equation}
where $i$ and $j$ are indices representing the spatial and/or Faraday components, and $S_{\rm 1.4\,GHz}$ is the flux density of the corresponding spatial component at 1.4\,GHz (listed in Paper II). This formulation is a modified version of that from \cite{osullivan17}, where the FDs were weighted by $p$ instead. The uncertainties in $\overline\phi$ are again propagated by Monte Carlo simulations as above. We will compare the $\overline\phi$ values against the \citetalias{taylor09} RM values to determine whether the source suffers from $n\pi$-ambiguity. The results are listed in Table~\ref{table:rmsyn}. 

We find that two of our target sources (J084600$-$261054 and J234033$+$133300) are unpolarised (less than the $6\sigma$ cutoff levels at $0.07$ and $0.06$ per cent, respectively), and therefore are excluded in the subsequent stages of our study in this paper\footnote{We believe this discrepancy with \citetalias{taylor09} in polarisation level is due to the off-axis polarisation leakage in the NVSS data, and we shall investigate this in detail in the companion Paper II.}. Furthermore, the spatial double J162706$-$091705 hosts one unpolarised component (a) and a polarised component (b). Out of the remaining 21 sources (five being spatial doubles) with reliable polarisation signals, nine have $\overline\phi$ disagreeing with the \citetalias{taylor09} RM values by about $\pm 652.9\,{\rm rad\,m}^{-2}$, and 11 have the two sets of values agreeing within $60\,{\rm rad\,m}^{-2}$. The only remaining source J154936$+$183500 is a special case, with $\overline\phi$ and ${\rm RM}_{\rm TSS09}$ values deviating by $307.5\,{\rm rad\,m}^{-2}$ (see Section~\ref{sec:j1549} for discussion on this source).

We further performed a per-pixel RM-Synthesis analysis to the extended sources J094750$-$371528 and J224549$+$394122 (also known as 3C 452), presented in Appendix~\ref{sec:spatial}. This allows the Faraday complexities of these two sources to be resolved spatially, leading to interesting comparisons with the RM-Synthesis results above and \textit{QU}-fitting results in Section~\ref{sec:qu}.

\subsection{Stokes \textit{QU}-fitting \label{sec:qu}}
We complement our RM-Synthesis results in Section~\ref{sec:rmsyn} with Stokes \textit{QU}-fitting analysis \citep[e.g.][]{farnsworth11,osullivan12}. Tests with synthetic data have shown that \textit{QU}-fitting can perform better than RM-Synthesis for sources composed of two Faraday thin components \citep{sun15}. The main difference between these two techniques is that the former is non-parametric, while the latter requires input astrophysical models. These models consist of one or more Faraday components added together, which can correspond to discrete astrophysical sources or emitting volumes with different physical parameters within our telescope beam or flux integration region. For our study, we considered the following Faraday components \citep{burn66,sokoloff98,osullivan12}:
\begin{enumerate}
\item \textit{Thin:} A purely synchrotron-emitting volume, with Faraday rotation occurring in a foreground screen with a homogeneous magnetic field and thermal electron density. The complex polarisation is given by
\begin{equation}
p_j(\lambda^2) = p_{0,j} e^{2i({\rm PA}_{0,j} + \phi_j \lambda^2)}{\rm .}
\end{equation}
\item \textit{Burn slab:} This depicts a volume that is simultaneously synchrotron-emitting and Faraday rotating, with no foreground Faraday rotating screens. The magnetic fields, thermal electron densities, and cosmic rays densities in the slab are all uniform. The complex polarisation is given by
\begin{equation}
p_j(\lambda^2) = p_{0,j} \frac{\sin (\phi_j \lambda^2)}{\phi_j \lambda^2} e^{2i({\rm PA}_{0,j} + \frac{1}{2}\phi_j \lambda^2)}{\rm .}
\end{equation}
\item \textit{Burn slab with foreground screen:} This is the same as a Burn slab component, except there is a homogeneous foreground rotating screen giving rise to an extra FD of $\phi_{\rm fg}$. The complex polarisation is given by
\begin{equation}
p_j(\lambda^2) = p_{0,j} \frac{\sin (\phi_j \lambda^2)}{\phi_j \lambda^2} e^{2i[{\rm PA}_{0,j} + (\frac{1}{2}\phi_j+\phi_{\rm fg}) \lambda^2]}{\rm .}
\end{equation}
\item \textit{External Faraday dispersion:} In addition to the homogeneous Faraday screen for a thin component, an external turbulent Faraday screen lies in front of the synchrotron-emitting volume. This turbulent screen leads to a dispersion in FD ($\sigma_\phi$) through different lines of sight to the emitting volume (within the telescope beam or the flux integration region), causing depolarisation effects. The complex polarisation is given by
\begin{equation}
p_j(\lambda^2) = p_{0,j} e^{-2 \sigma_{\phi,j}^2 \lambda^4} e^{2i({\rm PA}_{0,j} + \phi_j \lambda^2)}{\rm .}
\end{equation}
\item \textit{Internal Faraday dispersion:} This is similar to the Burn slab above, except that in the simultaneously emitting and Faraday rotating volume there is also a turbulent magnetic field component. The complex polarisation is given by
\begin{equation}
p_j(\lambda^2) = p_{0, j} e^{2i{\rm PA}_{0, j}} \left( \frac{1 - e^{i \phi_j \lambda^2 - 2 \sigma_{\phi, j}^2 \lambda^4}}{2\sigma_{\phi, j}^2 \lambda^4 - i \phi_j \lambda^2} \right){\rm .}
\end{equation}
\item \textit{Internal Faraday dispersion with foreground screen:} This is the same as the internal Faraday dispersion component, but there is a homogeneous foreground rotating screen leading to an extra FD of $\phi_{\rm fg}$. The complex polarisation is given by
\begin{equation}
p_j(\lambda^2) = p_{0, j} e^{2i({\rm PA}_{0, j} + \phi_{\rm fg} \lambda^2)} \left( \frac{1 - e^{i \phi_j \lambda^2 - 2 \sigma_{\phi, j}^2 \lambda^4}}{2\sigma_{\phi, j}^2 \lambda^4 - i \phi_j \lambda^2} \right){\rm .}
\end{equation}
\end{enumerate}

\begin{table*}
\caption{Results of \textit{QU}-fitting on the New Broadband VLA Data \label{table:qufit}}
\begin{tabular}{lcccccccc}
\hline
\multicolumn{1}{c}{Source} & Faraday & $\phi$ & $p$ & ${\rm PA}_0$ & $\phi_{\rm fg}$ & $\sigma_\phi$ & $\chi_{\rm red}^2$ & $\Delta{\rm PI}/\overline{\rm PI}$$^a$ \\
\multicolumn{1}{c}{(NVSS)} & Model & (${\rm rad\,m}^{-2}$) & (\%) & ($^\circ$) & (${\rm rad\,m}^{-2}$) & (${\rm rad\,m}^{-2}$) && (\%) \\
\hline
\multicolumn{9}{c}{\textbf{Outliers (Reliable $\mathbf{RM}_\mathbf{TSS09}$)}} \\
\hline
J083930$-$240723 & 1T & $+325.9\pm 0.7\phantom{0}$ & $4.55\pm 0.05$ & $+4.6\pm 1.5$ & --- & --- & $1.7$ & $+4.5$\\
J084701$-$233701 & 1T & $+385.5\pm 1.5\phantom{0}$ & $3.03\pm 0.08$ & $+87.8\pm 3.5\phantom{0}$ & --- & --- & $2.1$ & $+1.2$\\
J090015$-$281758 & 1T & $+352.0\pm 0.3\phantom{0}$ & $4.42\pm 0.02$ & $+6.9\pm 0.6$ & --- & --- & $2.5$ & $+3.5$\\
J092410$-$290606a & 2T & $+526.3\pm 0.7\phantom{0}$ & $8.22\pm 0.08$ & $+60.7\pm 1.8\phantom{0}$ & --- & --- & $1.5$ & $+7.1$\\
 & & $+582.9\pm 12.2$ & $0.51\pm 0.08$ & $+69.4\pm 30.1$ & --- & --- & --- & --- \\
J092410$-$290606b & 1Ed & $+530.6\pm 0.6\phantom{0}$ & $4.13\pm 0.15$ & $+75.3\pm 1.4\phantom{0}$ & --- & $6.2\pm 0.9$ & $1.2$ & $+1.3$\\
J093349$-$302700 & 1B+fg & \phantom{0}$+24.9\pm 1.5\phantom{0}$ & $7.01\pm 0.15$ & $-32.9\pm 1.1\phantom{0}$ & $+328.8\pm 1.0$ & --- & $2.7$ & $+0.1$\\
J093544$-$322845a & 1T & $+393.5\pm 0.4\phantom{0}$ & $7.61\pm 0.07$ & $-49.5\pm 1.0\phantom{0}$ & --- & --- & $2.4$ & $+3.8$\\
J093544$-$322845b & 1T & $+387.4\pm 0.6\phantom{0}$ & $5.07\pm 0.06$ & $+74.0\pm 1.3\phantom{0}$ & --- & --- & $2.7$ & $+4.7$\\
J094750$-$371528$^\odot$ & 1B+fg & $+50.3\pm 1.5$ & $6.60\pm 0.31$ & $-8.6\pm 3.6$ & $+303.7\pm 2.0$ & --- & $1.5$ & $-27.4$\\
J162706$-$091705b & 2T & $-330.4\pm 1.7\phantom{0}$ & $12.38\pm 0.52\phantom{0}$ & $+8.1\pm 3.4$ & --- & --- & $1.1$ & $+10.9$\\
 & & $-376.2\pm 11.5$ & $1.58\pm 0.52$ & $-10.8\pm 21.3$ & --- & --- & --- & --- \\
J163927$-$124139a & 1T & $-328.3\pm 0.2\phantom{0}$ & $9.16\pm 0.04$ & $+31.8\pm 0.5\phantom{0}$ & --- & --- & $1.5$ & $+4.1$\\
J163927$-$124139b & 1B+fg & $+16.1\pm 1.3$ & $10.78\pm 0.14\phantom{0}$ & $+68.3\pm 0.6\phantom{0}$ & $-343.8\pm 0.7$ & --- & $1.1$ & $+2.3$\\
J220205$+$394913 & 1T & $-367.3\pm 0.4\phantom{0}$ & $8.30\pm 0.07$ & $+59.2\pm 1.1\phantom{0}$ & --- & --- & $1.6$ & $+5.7$\\
J220927$+$415834 & 1T & $-338.3\pm 0.2\phantom{0}$ & $6.88\pm 0.04$ & $-12.2\pm 0.5\phantom{0}$ & --- & --- & $0.9$ & $+4.8$\\
\hline
\multicolumn{9}{c}{\textbf{Out-\textit{liars} ($n\pi$-ambiguity)}} \\
\hline
J022915$+$085125 & 2T & $+70.3\pm 0.7$ & $2.58\pm 0.03$ & $+44.0\pm 1.5\phantom{0}$ & --- & --- & $1.8$ & $-191.5$\\
 & & $-44.2\pm 0.7$ & $2.57\pm 0.03$ & $+89.8\pm 1.5\phantom{0}$ & --- & --- & --- & --- \\
J091145$-$301305a & 1T & $+244.8\pm 1.0\phantom{0}$ & $8.35\pm 0.17$ & $-26.9\pm 2.2\phantom{0}$ & --- & --- & $1.9$ & $+5.3$\\
J091145$-$301305b & 1T & $+247.5\pm 0.3\phantom{0}$ & $17.02\pm 0.09\phantom{0}$ & $-15.5\pm 0.6\phantom{0}$ & --- & --- & $2.3$ & $+4.3$\\
J094808$-$344010 & 2T & $+365.9\pm 1.0\phantom{0}$ & $4.30\pm 0.07$ & $+51.3\pm 2.1\phantom{0}$ & --- & --- & $2.8$ & $-1.4$\\
 & & $+466.4\pm 5.0\phantom{0}$ & $0.84\pm 0.07$ & $-25.8\pm 10.9$ & --- & --- & --- & --- \\
J111857$+$123442 & 2T & $+99.4\pm 1.7$ & $0.66\pm 0.01$ & $+49.7\pm 3.7\phantom{0}$ & --- & --- & $3.7$ & $-4.9$\\
 & & $+11.2\pm 6.1$ & $0.20\pm 0.01$ & $+89.8\pm 13.1$ & --- & --- & --- & --- \\
J170934$-$172853 & 2B+fg & $+21.4\pm 1.1$ & $3.82\pm 0.06$ & $-31.6\pm 2.2\phantom{0}$ & $+107.5\pm 0.6$ & --- & $1.6$ & $+0.4$\\
 & & $-91.3\pm 4.3$ & $2.46\pm 0.18$ & $+32.4\pm 7.8\phantom{0}$ & $+107.5\pm 0.6$ & --- & --- & --- \\
J190255$+$315942 & 2T & $+168.0\pm 0.5\phantom{0}$ & $2.39\pm 0.02$ & $+30.1\pm 1.1\phantom{0}$ & --- & --- & $6.5$ & $+2.6$\\
 & & $+96.5\pm 0.9$ & $1.27\pm 0.02$ & $-51.7\pm 2.2\phantom{0}$ & --- & --- & --- & --- \\
J224412$+$405715 & 1Ed & $-320.8\pm 0.5\phantom{0}$ & $3.86\pm 0.08$ & $+85.3\pm 1.0\phantom{0}$ & --- & $5.7\pm 0.5$ & $1.4$ & $-2.8$\\
J224549$+$394122$^\odot$ & 1Id+fg & $-19.4\pm 4.4$ & $7.40\pm 0.04$ & $+7.2\pm 0.5$ & $-270.7\pm 0.8$ & $13.9\pm 0.3\phantom{0}$ & $3.8$ & $-2.7$\\
J235728$+$230226 & 2T & $+36.0\pm 7.6$ & $0.22\pm 0.02$ & $+54.3\pm 17.1$ & --- & --- & $1.4$ & $+18.5$\\
 & & $+118.4\pm 14.8$ & $0.11\pm 0.02$ & $-68.9\pm 33.8$ & --- & --- & --- & --- \\
\hline
\multicolumn{9}{c}{\textbf{Others}} \\
\hline
J084600$-$261054$^\times$ & --- & --- & --- & --- & --- & --- & --- & --- \\
J154936$+$183500$^?$ & 3T & $-318.4\pm 1.8\phantom{0}$ & $0.42\pm 0.02$ & $+30.1\pm 4.3\phantom{0}$ & --- & --- & $2.0$ & $-30.9$\\
 & & $+70.3\pm 3.0$ & $0.31\pm 0.02$ & $-27.3\pm 7.1\phantom{0}$ & --- & --- & --- & --- \\
 & & $-12.0\pm 4.2$ & $0.23\pm 0.02$ & $+34.8\pm 9.7\phantom{0}$ & --- & --- & --- & --- \\
J234033$+$133300$^\times$ & --- & --- & --- & --- & --- & --- & --- & --- \\
\hline
\multicolumn{9}{l}{\texttt{NOTE} --- Key to the Faraday components: T: Thin; B: Burn slab; B+fg: Burn slab with foreground screen; Ed: External Faraday} \\
\multicolumn{9}{l}{\phantom{\texttt{NOTE} --- }dispersion; Id: Internal Faraday dispersion; Id+fg: Internal Faraday dispersion with foreground screen} \\
\multicolumn{9}{l}{$^a$$\Delta{\rm PI} = {\rm PI}_1 - {\rm PI}_2$ and $\overline{\rm PI} = ({\rm PI}_1 + {\rm PI}_2)/2$, where ${\rm PI}_1$ and ${\rm PI}_2$ are the predicted polarised intensities at the two NVSS IFs according to} \\
\multicolumn{9}{l}{\phantom{$^a$}the best-fit model and fitted $\alpha_{\rm L}$ (reported in Paper II), without taking bandwidth depolarisation into account} \\
\multicolumn{9}{l}{$^\times$Unpolarised sources} \\
\multicolumn{9}{l}{$^?$Special case compared to \citetalias{taylor09} catalogue (see Section~\ref{sec:j1549})} \\
\multicolumn{9}{l}{$^\odot$Extended sources}
\end{tabular}
\end{table*}

A caveat of the \textit{QU}-fitting technique here is that, similar to RM-Synthesis in Section~\ref{sec:rmsyn}, the intrinsic polarisation fraction $p_{0,j}$ obtained from this analysis is the polarised intensity of the component $j$ divided by the total intensity of the entire spatial component, since this analysis does not separate the total intensity into corresponding Faraday components.

We deployed 10 different models to fit the observed $q$ and $u$ values of our target sources: single thin (1T), double thin (2T), triple thin (3T), single Burn slab (1B), double Burn slab (2B), single Burn slab with foreground screen (1B+fg), double Burn slab with foreground screen (2B+fg), single external Faraday dispersion (1Ed), single internal Faraday dispersion (1Id), and single internal Faraday dispersion with foreground screen (1Id+fg). The complex polarisation of the models are constructed by adding together that of the constituent Faraday components [$p(\lambda^2) = \sum_j p_j(\lambda^2)$]. In other words, the Faraday components of each model are assumed to be spatially distributed perpendicular to the line of sight. For the double Burn slab with foreground screen model, both of the Burn slab components are subjected to the same foreground FD, instead of having individual $\phi_{\rm fg}$ values. The best-fit parameters and their uncertainties of each of the models for each target source are obtained, along with the reduced chi squared values ($\chi_{\rm red}^2$) and the Bayesian information criterion \citep[BIC; e.g.][]{osullivan12,schnitzeler19}. We rejected models where the $p_{0,j}$ and/or $\sigma_{\phi,j}$ values are less than two times of the uncertainties. The remaining models for each source are ranked according to the BIC values (with a lower value signifying a better model), and the best for each source is listed in Table~\ref{table:qufit} and plotted in Figure~\ref{fig:qufit} in the Online Supplementary Materials (Appendix~\ref{sec:online}).

\section{Discussion} \label{sec:discussion}
\subsection{The $n\pi$-ambiguity in the NVSS RM Catalogue}
\begin{figure*}
\includegraphics[width=\doublecolumnwidth]{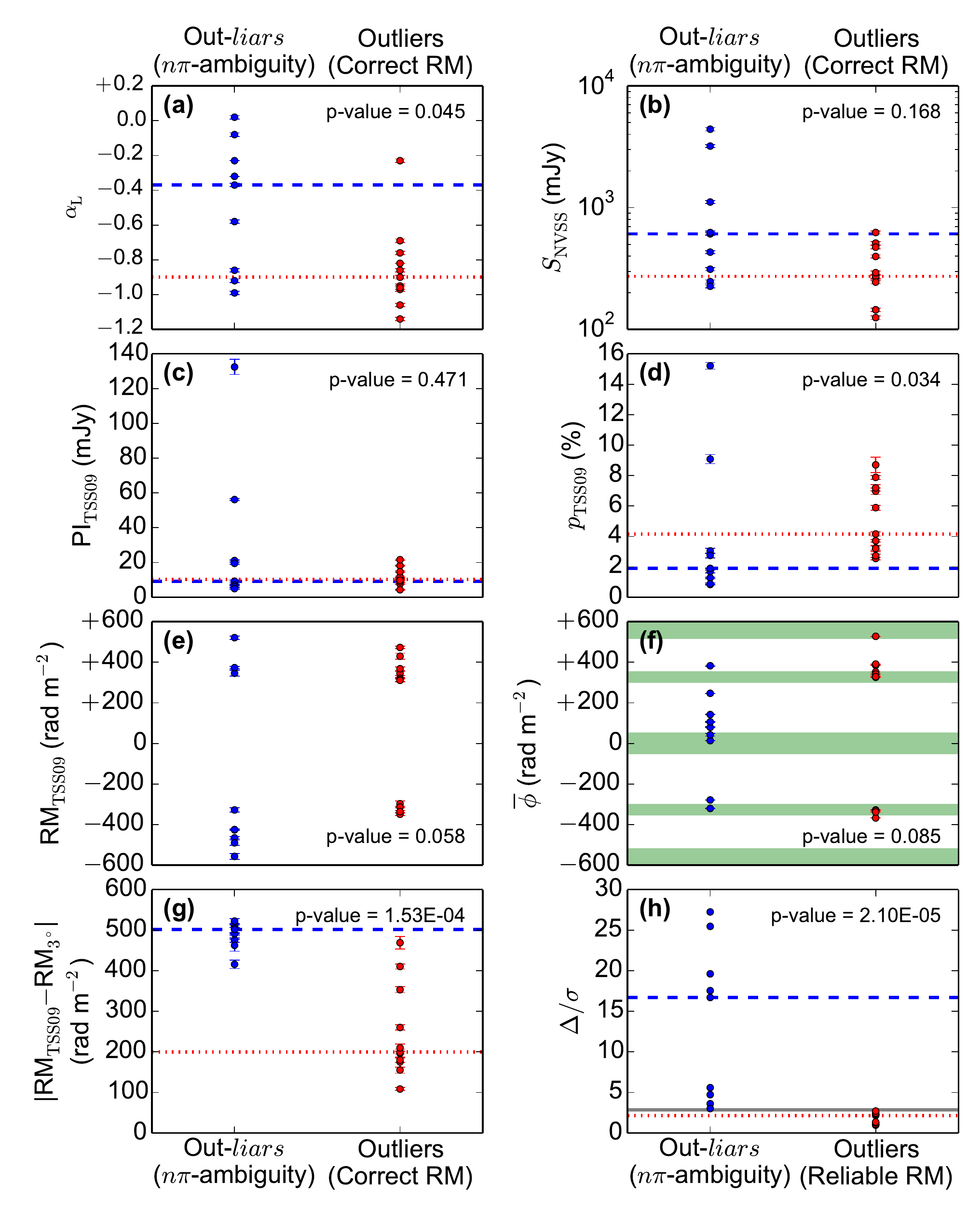}
\caption{Select parameters of our 20 target sources separated into out-\textit{liars} (left; in blue) and outliers (right; in red). The p-value from two-sample KS-test is reported in each panel. In relevant cases, the medians of the two populations are plotted as blue dashed (for out-\textit{liars}) and red dotted (for outliers) lines. In panel (f), the areas highlighted in green corresponds to the $|{\rm RM}|$ ranges of $< 50$, $> 520$, and $\approx 326.5\,{\rm rad\,m}^{-2}$, within which the $R_0$ parameter has limited reliability \citepalias{taylor09}. The grey solid line in panel (h) indicates the cutoff level at 2.85 (see text). \label{fig:npi_diag}}
\end{figure*}

In Section~\ref{sec:rmsyn}, we compared our $\overline\phi$ values from RM-Synthesis performed on the new broadband data with narrowband RMs from the NVSS RM catalogue \citepalias{taylor09}. Nine out of 21 of our polarised target sources (43 per cent) have $\overline\phi$ values deviating by approximately $\pm 652.9\,{\rm rad\,m}^{-2}$ from the corresponding ${\rm RM}_{\rm TSS09}$. The discrepancy is almost certainly due to $n\pi$-ambiguity in the \citetalias{taylor09} catalogue. In an attempt to unveil the cause(s) and possible diagnostic(s) of this, we divided our sources into the two classes -- out-\textit{liars} and outliers -- and compared select observed quantities. Specifically, we investigated the distributions of spectral index from our L-band observations ($\alpha_{\rm L}$; reported in Paper II), NVSS flux density ($S_{\rm NVSS}$), \citetalias{taylor09} polarised intensity (${\rm PI}_{\rm TSS09}$), \citetalias{taylor09} polarisation fraction ($p_{\rm TSS09}$), ${\rm RM}_{\rm TSS09}$, $\overline\phi$, $|{\rm RM}_{\rm TSS09} - {\rm RM}_{3^\circ}|$, and $|{\rm RM}_{\rm TSS09} - {\rm RM}_{3^\circ}|/\sigma_{3^\circ}$, with ${\rm RM}_{3^\circ}$ and $\sigma_{3^\circ}$ defined below. For each parameter, we performed two-sample Kolmogorov--Smirnov test (KS-test) with the null hypothesis being that the two samples are drawn from the same population. The above parameters are plotted in Figure~\ref{fig:npi_diag}, with their corresponding KS-test p-values also reported. We adopted the standard p-value cutoff of 0.05 (a larger p-value favours the null hypothesis), and concluded that the two populations have different distributions in $\alpha_{\rm L}$, $p_{\rm TSS09}$, $|{\rm RM}_{\rm TSS09} - {\rm RM}_{3^\circ}|$, and $|{\rm RM}_{\rm TSS09} - {\rm RM}_{3^\circ}|/\sigma_{3^\circ}$, which we will discuss in detail below. On the other hand, our KS-test results suggest that the two classes of sources likely originate from the same population in $S_{\rm NVSS}$, ${\rm PI}_{\rm TSS09}$, ${\rm RM}_{\rm TSS09}$, and $\overline\phi$, with p-values of 0.168, 0.471, 0.058, and 0.085, respectively. However, note that given this small sample size (nine and 11 in the two classes), we cannot rule out the possibility that our statistical analysis here could be biased by random statistical anomalies. Below, we will also explore the effects of FD ranges and Faraday complexities (Table~\ref{table:complex}) on $n\pi$-ambiguity in \citetalias{taylor09} catalogue, and investigate the special case J154936$+$183500, which has a difference between $\overline\phi$ and ${\rm RM}_{\rm TSS09}$ consistent with neither the outlier nor the out-\textit{liar} cases.

\begin{figure}
\includegraphics[width=\columnwidth]{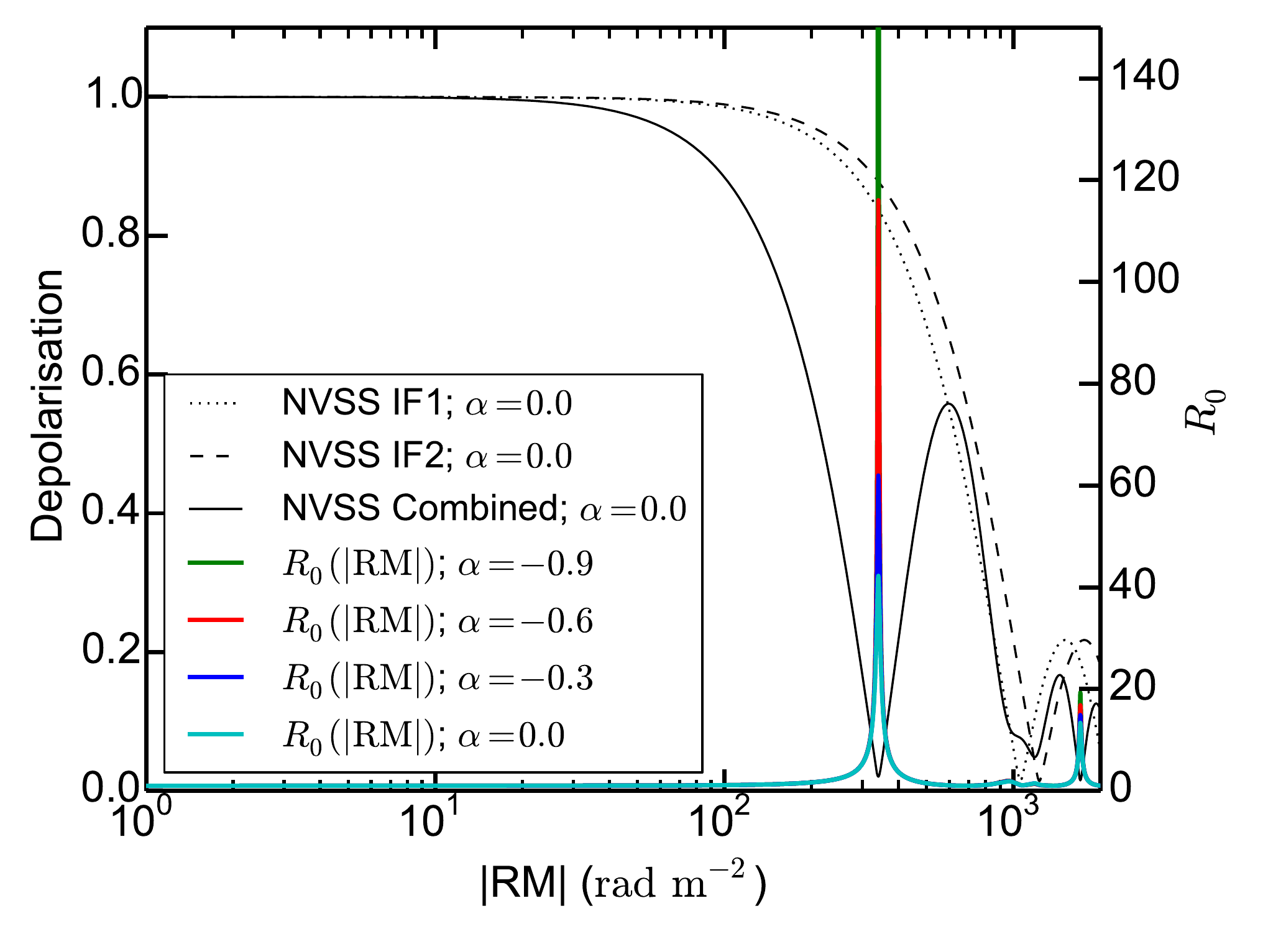}
\caption{Bandwidth depolarisation and $R_0 = ({\rm PI}_1+{\rm PI}_2)/2{\rm PI}_c$ of the NVSS observational setup, assuming a Faraday simple source with constant $p(\lambda^2)$. The bandwidth depolarisation of the two NVSS IFs, as well as that combining both bands, are shown as the black curves for the case with spectral index of $\alpha = 0.0$. $R_0$ with different $\alpha$ values are plotted as the coloured curves, with the $y$-axis truncated at $R_0 = 150$ because the peak of $R_0$ for $\alpha = -0.9$ at ${\rm RM} \approx 340\,{\rm rad\,m}^{-2}$ reaches about 900, which can have obscured the other lower peaks. \label{fig:r0_rm}}
\end{figure}

\subsubsection{The \citetalias{taylor09} $n\pi$-ambiguity Rejection Algorithm \label{sec:t09review}}
Before looking into the dependence of $n\pi$-ambiguity on various parameters, we review the algorithm devised by \citetalias{taylor09} to minimise $n\pi$-ambiguity in their catalogue. This algorithm picks the most probable RM value for each source based on the following three constraints. First, they assumed that at most only a single PA wrap can occur between the two NVSS IFs. This imposes a limit of $|{\rm RM}_{\rm TSS09}| \leq 1306\,{\rm rad\,m}^{-2}$ for all sources. Second, they introduced the parameter 
\begin{equation}
R_0 = \frac{{\rm PI}_1+{\rm PI}_2}{2{\rm PI}_c}{\rm ,}\label{eq:r0}
\end{equation}
where ${\rm PI}_1$, ${\rm PI}_2$, and ${\rm PI}_c$ are the polarised intensities in NVSS IF1, IF2, and combined band, respectively. Since the measured PI is a function of the source $|{\rm RM}|$ due to bandwidth depolarisation in the NVSS observational setup, the parameter $R_0$ in turn is also a function of $|{\rm RM}|$ (Figure~\ref{fig:r0_rm}). \citetalias{taylor09} compared the observed $R_0$ of each source with the predicted $R_0$ values at the few possible RM values, with a likelihood assigned to each possible RM. This means \citetalias{taylor09} assumed that the differences among ${\rm PI}_1$, ${\rm PI}_2$, and ${\rm PI}_c$ are only due to bandwidth depolarisation, but not caused by other effects such as spectral indices and Faraday complexities. Lastly, they rejected candidate RM values that deviated significantly from the RM values of surrounding sources. Specifically, for each source they computed the median ${\rm RM}_{\rm TSS09}$ of neighbouring sources within a radius of $3^\circ$, and only accepted candidate RM values within $520\,{\rm rad\,m}^{-2}$ from the median RM. This implicitly assumes that the RM of individual sources cannot deviate significantly from that of their neighbours due to intrinsic RM or spatial fluctuations of foreground RM. The most likely candidate RM remaining is then reported as the ${\rm RM}_{\rm TSS09}$.

\subsubsection{Dependence on Spectral Index \label{sec:npi_alpha}}
As seen in panel (a) of Figure~\ref{fig:npi_diag}, the out-\textit{liars} and outliers appear to exhibit different distributions in $\alpha_{\rm L}$. This is further supported by the KS-test p-value of 0.045. The out-\textit{liars} have $\alpha_{\rm L}$ spread evenly over a wide range from $-0.99$ to $+0.02$, with a median of $-0.37$, while the $\alpha_{\rm L}$ of most (ten out of 11) of the outliers cluster between $-1.14$ and $-0.69$, with a median of $-0.9$. It would be natural to directly link this discrepancy to the $R_0$ parameter used in the \citetalias{taylor09} algorithm. This is because the change in PI across $\lambda^2$ caused by spectral index effects could be mistaken as bandwidth depolarisation by the $R_0$ algorithm, possibly leading to $n\pi$-ambiguity. To test this hypothesis, we simulated $R_0$ as a function of $|{\rm RM}|$ for several different $\alpha_{\rm L}$ values ($0.0$, $-0.3$, $-0.6$, and $-0.9$), assuming Faraday simplicity. The results are shown in Figure~\ref{fig:r0_rm}. We find that the $R_0$ values at any given $|{\rm RM}|$ are only weakly dependent on $\alpha_{\rm L}$ except near the peak at $|{\rm RM}| \approx 340\,{\rm rad\,m}^{-2}$ where the predicted $R_0$ diverge. This means that $R_0$ could be less effective in distinguishing different RM values for sources with true $|{\rm RM}| \approx 340\,{\rm rad\,m}^{-2}$. \citetalias{taylor09} also reached similar conclusion regarding this $|{\rm RM}|$ range but for different reasons (see Section~\ref{sec:rmrange}). However, only two out of nine out-\textit{liars} (J094808$-$344010 and J224412$+$405715) reside in this $|{\rm RM}|$ range, and therefore spectral index dependence cannot explain most of our $n\pi$-ambiguity sources.

\subsubsection{Dependence on Polarised Intensity and Polarisation Fraction}
We show in panel (c) of Figure~\ref{fig:npi_diag} the distribution of the two classes in ${\rm PI}_{\rm TSS09}$, which is the average PI in the two IFs [i.e.\ $({\rm PI}_1 + {\rm PI}_2)/2$]. The KS-test p-value of 0.471 suggests that the two samples have the same underlying distribution in ${\rm PI}_{\rm TSS09}$. It is worth noting, however, that while a previous study of 37 radio sources with high ${\rm PI}_{\rm TSS09}$ ($ > 200\,{\rm mJy}$) also at 1--2\,GHz with the Allen Telescope Array (ATA) found no $n\pi$-ambiguity in the \citetalias{taylor09} catalogue, we have identified J224549$+$394122, which is spatially extended with ${\rm PI} \approx 557.8 \pm 1.1\,{\rm mJy}$ in our observation and ${\rm PI}_{\rm TSS09} = 132.4 \pm 4.3\,{\rm mJy}$ (the difference can be due to how \citetalias{taylor09} extracted Stokes \textit{Q} and \textit{U} values for spatially resolved sources), as an out-\textit{liar}. This shows that not all sources with high PI have reliable \citetalias{taylor09} RM values.

The discrepancies between the two populations in the fractional polarisation reported in the \citetalias{taylor09} catalogue ($p_{\rm TSS09}$) is more apparent, with KS-test p-value of 0.034. The $p_{\rm TSS09}$ values are plotted in panel (d) of Figure~\ref{fig:npi_diag}. Most (79 per cent) out-\textit{liars} are concentrated at 0.8--3.1 per cent, while the outliers spread more evenly between 2.6 and 8.7 per cent. The median $p_{\rm TSS09}$ of out-\textit{liars} and outliers are respectively 1.9 and 4.2 per cent. As we show in Paper II, sources with lower fractional polarisation are more susceptible to instrumental effects, particularly off-axis polarisation leakage, which can diminish the effectiveness of the \citetalias{taylor09} algorithm. However, this alone cannot explain all the out-\textit{liars} we identified, since two of them are highly polarised at $9.1 \pm 0.3$ (J224549$+$394122) and $15.2 \pm 0.2$ per cent (J091145$-$301305). 

\subsubsection{Dependence on FD and RM Ranges \label{sec:rmrange}}
\citetalias{taylor09} pointed out that $R_0$ (Equation~\ref{eq:r0}) could be less effective in selecting the correct RM values for sources with true $|$RM$|$ falling within ranges of $< 50\,{\rm rad\,m}^{-2}$, $>520\,{\rm rad\,m}^{-2}$, and $\approx 326.5\,{\rm rad\,m}^{-2}$ (taken as $301$--$352\,{\rm rad\,m}^{-2}$ here). For the case of $<50$ and $>520\,{\rm rad\,m}^{-2}$, that is because for both cases $R_0 \approx 1$, making it difficult to discern the correct RM value. The $|{\rm RM}|$ value of $\approx 326.5\,{\rm rad\,m}^{-2}$ is also believed to be a challenge for the \citetalias{taylor09} algorithm, since (1) there is almost complete bandwidth depolarisation for the combined band, and (2) $R_0$ cannot distinguish between the case of $+326.5$ and $-326.5\,{\rm rad\,m}^{-2}$ as the predicted $R_0$ are the same.

We plotted $\overline\phi$ of the two samples in panel (f) of Figure~\ref{fig:npi_diag}, with the above ranges shaded in green. Out of our sample, ten sources fall into these ranges, with only three being out-\textit{liars} -- J022915$+$085125 with $+13.6\pm 1.0\,{\rm rad\,m}^{-2}$, J224412$+$405715 with $-320.4 \pm 0.6\,{\rm rad\,m}^{-2}$, and J235728$+$230226 with $+42.3 \pm 6.1\,{\rm rad\,m}^{-2}$. It is apparent that out-\textit{liars} do not preferentially fall into the above RM ranges, and our samples within those ranges are more likely to have correct ${\rm RM}_{\rm TSS09}$ than suffer from $n\pi$-ambiguity.

\subsubsection{Dependence on Faraday Complexity}
Faraday complexity (formally defined in Section~\ref{sec:complex_def}) could be one of the reasons for the presence of $n\pi$-ambiguity in the NVSS RM catalogue. As summarised in Table~\ref{table:complex}, six out of the nine (67 per cent) out-\textit{liars} are Faraday complex from our RM-Synthesis results, while only one out of the 11 (9 per cent) outliers show Faraday complexities from the same analysis. We can draw similar conclusion from the \textit{QU}-fitting results, with eight out of nine (89 per cent) and five out of 11 (45 per cent) sources being Faraday complex, respectively. This may be because the $R_0$ algorithm can be affected by both non-linear PA and varying $p$ across $\lambda^2$. 

\begin{table}
\caption{Number of Faraday Simple/Complex Sources \label{table:complex}}
\begin{tabular}{ccc|cc}
\hline
& \multicolumn{2}{c|}{RM-Synthesis} & \multicolumn{2}{c}{\textit{QU}-fitting} \\
& Faraday & Faraday & Faraday & Faraday \\
& Simple & Complex & Simple & Complex \\
\hline
Out-\textit{liars} & 3 & 6 & 1 & 8 \\
($n\pi$-ambiguity) & & & & \\
Outliers & 10 & 1 & 6 & 5 \\
(Reliable ${\rm RM}_{\rm TSS09}$) & & & & \\
\hline
\end{tabular}
\end{table}

We test the possibility of the latter by quantifying the amount of Faraday depolarisation due to Faraday complexities. For each source, we adopted the best-fit model from \textit{QU}-fitting, as well as spectral index $\alpha_{\rm L}$ from Paper II, to compute the ${\rm PI}$ at the two NVSS IFs (${\rm PI}_1$ and ${\rm PI}_2$) \emph{without} taking bandwidth depolarisation into account. A depolarisation parameter is defined as
\begin{equation}
\frac{\Delta{\rm PI}}{\overline{\rm PI}} = \frac{{\rm PI}_1 - {\rm PI}_2}{({\rm PI}_1 + {\rm PI}_2)/2}{\rm ,}
\end{equation}
which is listed in Table~\ref{table:qufit} for each source. Note that the values for even sources best characterised by the single thin model are non-zero because of the effect of the spectral index, which leads to a positive $\Delta{\rm PI}/\overline{\rm PI}$ with negative $\alpha_{\rm L}$. Apart from J022915$+$085125 which has a large $|\Delta{\rm PI}/\overline{\rm PI}| = 191.5$ because $p(\lambda^2)$ at NVSS IF1 approaches zero, we do not see clear signs of out-\textit{liars} having larger $|\Delta{\rm PI}/\overline{\rm PI}|$, as would be expected if the Faraday depolarisation affects the $R_0$ algorithm leading to $n\pi$-ambiguity.

\subsubsection{Dependence on RM of Neighbouring Sources \label{sec:neighbour}}
For each of our target sources, we evaluated the medians (${\rm RM}_{3^\circ}$) and standard deviations ($\sigma_{3^\circ}$) of ${\rm RM}_{\rm TSS09}$ values of the neighbouring sources within a radius of $3^\circ$. These values are listed in column 4 of Table~\ref{table:fg}. On average, there are 28 neighbours to our target sources within the $3^\circ$ radius circles. Assuming that ${\rm RM}_{\rm TSS09}$ values are correct for most of the neighbouring sources, ${\rm RM}_{3^\circ}$ and $\sigma_{3^\circ}$ would respectively represent the RM contribution by large-scale Galactic and/or intergalactic component(s), and the spatial fluctuations of the above-mentioned foreground RM components superimposed on the statistical spread of intrinsic RM of the neighbouring sources.

The out-\textit{liars} clearly deviate from the outliers in $|{\rm RM}_{\rm TSS09} - {\rm RM}_{3^\circ}|$ values, as we show in panel (g) of Figure~\ref{fig:npi_diag}. The out-\textit{liars} gather within $416$--$522\,{\rm rad\,m}^{-2}$, with a median of $502\,{\rm rad\,m}^{-2}$, while the outliers spread through $109$ to $469\,{\rm rad\,m}^{-2}$, with a median of $199\,{\rm rad\,m}^{-2}$. The large values for out-\textit{liars} are clearly due to $n\pi$-ambiguity, leading to discrepancies between the individual ${\rm RM}_{\rm TSS09}$ and the respective ${\rm RM}_{3^\circ}$. Large values are also found for three outliers, which could stem from spatial variations of the foreground RM structures around the positions of those of our targets. If this is the case, we would expect high $\sigma_{3^\circ}$ values from those outliers as well.

An even clearer diagnostic is therefore the deviation in RM in units of $\sigma_{3^\circ}$. We plotted this ($\left| {\rm RM}_{\rm TSS09} - {\rm RM}_{3^\circ}\right|/\sigma_{3^\circ}$; shortened as $\Delta/\sigma$ in text below) in panel (h) of Figure~\ref{fig:npi_diag}. Indeed, we found that all outliers converged to $0.97$--$2.71$ in $\Delta/\sigma$, meaning that those with high $|{\rm RM}_{\rm TSS09} - {\rm RM}_{3^\circ}|$ also have high $\sigma_{3^\circ}$, matching our expectation above. The out-\textit{liars}, on the other hand, have $\Delta/\sigma$ spread over $3.02$ to $27.24$, since the large $|{\rm RM}_{\rm TSS09} - {\rm RM}_{3^\circ}|$ are due to $n\pi$-ambiguity and not necessarily accompanied by high $\sigma_{3^\circ}$ due to spatial variations of the foreground. There is an apparent cutoff between the two populations at about $2.85$. Since this $\Delta/\sigma$ parameter can be computed from the listed information from the NVSS RM catalogue without any extra information, this could be useful for identification of $n\pi$-ambiguity sources in the \citetalias{taylor09} catalogue (see Section~\ref{sec:howmany}).

\subsubsection{How Many \citetalias{taylor09} Sources Suffer from $n\pi$-ambiguity? \label{sec:howmany}}

We apply our findings from Section~\ref{sec:neighbour} to estimate how many \citetalias{taylor09} sources suffer from $n\pi$-ambiguity. The $\Delta/\sigma$ values for all of the 37,543 \citetalias{taylor09} sources have been computed. There is an average of 33 neighbouring sources for each \citetalias{taylor09} source. Sources with $\Delta/\sigma$ larger than $2.70$, $2.85$, and $3.00$ are identified, corresponding to loose, moderate, and strict cutoffs respectively according to Section~\ref{sec:neighbour}. Although we found that 837, 701, and 603 sources satisfy the above lower limits in $\Delta/\sigma$ respectively, we also noted that some of such sources have low $|{\rm RM}_{\rm TSS09} - {\rm RM}_{3^\circ}|$. These sources may be located at regions with smooth RM foreground leading to low $\sigma_{3^\circ}$ and high $\Delta/\sigma$, but not suffering from $n\pi$-ambiguity. We therefore imposed another constraint of $|{\rm RM}_{\rm TSS09} - {\rm RM}_{3^\circ}| \geq 400\,{\rm rad\,m}^{-2}$. This results in 56, 49, and 48 $n\pi$-ambiguity candidates in the entire \citetalias{taylor09} catalogue, depending on whether we adopt the loose, moderate, or strict cutoffs as defined above, respectively. Note that this is a lower limit estimated by the $\Delta/\sigma$ parameter only, which may not exhaust the entire $n\pi$-ambiguity population of \citetalias{taylor09} (see below). On the other hand, EGSs with high intrinsic FD (or RM) magnitudes of $\gtrsim 400\,{\rm rad\,m}^{-2}$ might also be included under the above selection criteria.

We further compared our list of $n\pi$-ambiguity candidates with the literature to verify the accuracy of our $\Delta/\sigma$ criterion. The wrongly classified sources (if any) can be separated into two categories -- false-positives ($n\pi$-ambiguity candidates that actually have reliable ${\rm RM}_{\rm TSS09}$) and false-negatives (sources that actually suffer from $n\pi$-ambiguity but not picked up by our algorithm above). No false-positives have been identified after consulting catalogues of polarised sources verified to have reliable ${\rm RM}_{\rm TSS09}$ \citep{mao10,law11,vaneck11,mao12,rawes18,betti19}, suggesting that our list of $n\pi$-ambiguity candidates is accurate. We further compare our findings with the known \citetalias{taylor09} $n\pi$-ambiguity sources listed in the literature to look for the false-negatives. \cite{vaneck11} reported RM values of 194 EGSs on the Galactic plane ($|b| \leq 5^\circ$) with their observations, of which 146 were cross-matched with \citetalias{taylor09}. From this sample, 13 sources (9\,\%) were found to suffer from $n\pi$-ambiguity in \citetalias{taylor09}. Most of these 13 sources are concentrated in the inner Galaxy ($35^\circ \leq l \leq 52^\circ$), with 11 out of the 15 cross-matches in that region suffering from $n\pi$-ambiguity. This is likely linked to the complex large-scale magnetic field structure of the Milky Way manifested as large $|{\rm RM}|$ and rapid changes in RM in small spatial scales of a few degrees \citep[e.g.,][]{sun08,vaneck11,jansson12}, ultimately leading to the concentration of $n\pi$-ambiguity sources there. However, using our $\Delta/\sigma$ parameter defined above only one out of those 11 $n\pi$-ambiguity sources found there is correctly classified as an $n\pi$-ambiguity candidate. This means that our $n\pi$-ambiguity candidates list from $\Delta/\sigma$ is conservative, i.e.\ there can be more than 50 $n\pi$-ambiguity sources in the entire \citetalias{taylor09} catalogue.

\subsubsection{NVSS J154936$+$183500: A Special Case \label{sec:j1549}}
Upon comparison between our broadband $\overline\phi$ with narrowband ${\rm RM}_{\rm TSS09}$ (Section~\ref{sec:rmsyn}), we identified J154936$+$183500 which has the two values differing by $307.5\,{\rm rad\,m}^{-2}$. This source can neither be classified as an out-\textit{liar} nor an outlier, as these two classes of sources should have deviating $\overline\phi$ and ${\rm RM}_{\rm TSS09}$ by about $652.9$ and $0\,{\rm rad\,m}^{-2}$ respectively. To rule out the possibility that this discrepancy of $307.5\,{\rm rad\,m}^{-2}$ is due to RM time variabilities, we compared the ${\rm RM}_{\rm TSS09}$ of this source with ${\rm RM}_{\rm VLA}$ from our Paper II. This ${\rm RM}_{\rm VLA}$ is obtained from our new observations within the NVSS frequency ranges only. We find that this source has ${\rm RM}_{\rm VLA} = -473.5 \pm 14.4\,{\rm rad\,m}^{-2}$, similar to its ${\rm RM}_{\rm TSS09} = -426.8 \pm 14.6\,{\rm rad\,m}^{-2}$. In other words, the difference of $307.5\,{\rm rad\,m}^{-2}$ above cannot be attributed to time variabilities.

This peculiar difference in $\overline\phi$ versus ${\rm RM}_{\rm TSS09}$ is likely due to its Faraday complexity. Both RM-Synthesis and \textit{QU}-fitting suggest that this source contains three Faraday components at FDs of about $-315$, $-20$, and $+75\,{\rm rad\,m}^{-2}$, with $p$ of about $0.46$, 0.27, and 0.33 per cent respectively. Such a wide spread of Faraday components over FD, combined with their similar fractional polarisation, results in highly non-linear PA across $\lambda^2$ in the NVSS bands, as well as in our broadband L-band. This leads to a poor agreement between $\overline\phi$ and ${\rm RM}_{\rm TSS09}$.

J154936$+$183500 is an example of sources that might not be suitable for RM grid experiments \citep[e.g.][]{gaensler05,mao10,vaneck11}. This is because of the large spread in FD of the three Faraday components within the telescope beam, suggesting that this source has large intrinsic FD ($\sim 100\,{\rm rad\,m}^{-2}$). For such case, narrowband RM values are clearly poor representations of the foreground magneto-ionic media, while techniques applied to broadband data such as extraction of absolute maxima in Faraday spectra \citep[e.g.][]{mao10,betti19} and using polarisation-weighted FD \citep[e.g.][and this work]{osullivan17} also may not give satisfactory results. This highlights the power of broadband spectro-polarimetric observations, which have opened up the possibility to identify such sources for careful treatments in RM grid experiments and/or further studies of their intrinsic polarisation properties.

\subsubsection{Summary of this Section}
In this Section, we showed the differences in the statistical distributions for several parameters of out-\textit{liars} ($n\pi$-ambiguity sources in \citetalias{taylor09}) versus outliers (sources with reliable ${\rm RM}_{\rm TSS09}$). We suggest that low $p_{\rm TSS09}$ could cause $n\pi$-ambiguity in \citetalias{taylor09} values. Also, out-\textit{liars} are found to have larger spread in $\alpha_{\rm L}$ and tends to be Faraday complex, while outliers are concentrated at steeper $\alpha_{\rm L}$ and are more likely Faraday simple. However, there may not be a direct relationship between these and $n\pi$-ambiguity. Out-\textit{liars} do not appear to preferentially fall within $|{\rm FD}|$ ranges of $< 50$, $> 520$, and $\approx 326.5\,{\rm rad\,m}^{-2}$. We further compared, for each of our target sources, their ${\rm RM}_{\rm TSS09}$ with the median (${\rm RM}_{3^\circ}$) and standard deviation ($\sigma_{3^\circ}$) of ${\rm RM}_{\rm TSS09}$ of neighbouring sources within a radius of $3^\circ$. All out-\textit{liars} cluster at $|{\rm RM}_{\rm TSS09} - {\rm RM}_{3^\circ}| \approx 500\,{\rm rad\,m}^{-2}$, while outliers span a range between $110$ to $470\,{\rm rad\,m}^{-2}$. Most interestingly, we found that $\Delta/\sigma = |{\rm RM}_{\rm TSS09} - {\rm RM}_{3^\circ}|/\sigma_{3^\circ}$ is an excellent diagnostic for $n\pi$-ambiguity in \citetalias{taylor09} catalogue. This parameter is an indicator of how much the ${\rm RM}_{\rm TSS09}$ value of each source deviates from the RM caused by foreground structures, in units of how much such foreground structures fluctuate spatially. There is a cutoff at $\sim 2.85$ between the two classes of sources, with out-\textit{liars} being above this cutoff and outliers below. Using this $\Delta/\sigma$ parameter, combined with a further constraint of $|{\rm RM}_{\rm TSS09} - {\rm RM}_{3^\circ}| \gtrsim 400\,{\rm rad\,m}^{-2}$ to discard sources situated behind smooth RM foregrounds with low $\sigma_{3^\circ}$, we estimate that at least 50 out of the 37,543 \citetalias{taylor09} sources can be affected by the $n\pi$-ambiguity effect.

\begin{table*}
\caption{Foreground Diagnostics to Our Target Sources \label{table:fg}}
\begin{tabular}{lcccccc}
\hline
\multicolumn{1}{c}{Source} & $\overline\phi$ & RM$_{\rm TSS09}$ & RM$_{3^\circ} \pm \sigma_{3^\circ}$$^a$ & RM$_{\rm O15}$$^b$ & $N_{\mathrm{H}\textsc{i}}$$^c$ & $I_{{\rm H}\alpha}$$^d$ \\
\multicolumn{1}{c}{(NVSS)} & (${\rm rad\,m}^{-2}$) & (${\rm rad\,m}^{-2}$) & (${\rm rad\,m}^{-2}$) & (${\rm rad\,m}^{-2}$) & ($10^{20}{\rm cm}^{-2}$) & (Rayleighs) \\
\hline
J022915$+$085125$^\dagger$ & \phantom{0}$+13.6 \pm 1.0$ & $+521.2 \pm 8.0$\phantom{0} & \phantom{0}$+0.9 \pm 19.1$ & \phantom{0}$+0.2 \pm 7.8$ & $6.15$$^E$ & $0.78$ \\
J083930$-$240723 & $+325.9 \pm 1.0$ & $+345.2 \pm 10.5$ & $+148.8 \pm 191.0$ & $+123.4 \pm 32.3$ & $7.05$$^G$ & $45.07$ \\
J084600$-$261054$^\times$ & --- & $+481.0 \pm 11.4$ & $+168.9 \pm 169.2$ & $+202.2 \pm 21.8$ & $7.67$$^G$ & $48.05$ \\
J084701$-$233701 & $+384.8 \pm 2.0$ & $+429.5 \pm 15.3$ & \phantom{0}$-39.5 \pm 216.2$ & \phantom{0}$+67.4 \pm 29.1$ & $8.17$$^G$ & $49.93$ \\
J090015$-$281758 & $+352.1 \pm 0.2$ & $+320.6 \pm 4.2$\phantom{0} & $+121.3 \pm 204.7$ & $+162.8 \pm 23.7$ & $12.2$$^G$ & $61.12$ \\
J091145$-$301305$^{\star\star\dagger}$ & $+246.9^{+0.3}_{-0.3}$ & $-426.1 \pm 3.5$\phantom{0} & \phantom{0}$+84.0 \pm 141.2$ & \phantom{0}$+81.5 \pm 28.6$ & $15.7$$^G$ & $33.24$ \\
J092410$-$290606$^{\star\star}$ & $+527.6^{+0.3}_{-0.3}$ & $+472.9 \pm 6.2$\phantom{0} & \phantom{0}$+62.5 \pm 183.9$ & \phantom{0}$+95.7 \pm 24.5$ & $8.48$$^G$ & $36.36$ \\
J093349$-$302700 & $+341.6 \pm 0.8$ & $+313.4 \pm 7.7$\phantom{0} & $+158.1 \pm 144.1$ & $+138.4 \pm 22.7$ & $9.64$$^G$ & $29.90$ \\
J093544$-$322845$^{\star\star}$ & $+390.9^{+0.3}_{-0.3}$ & $+368.1 \pm 9.3$\phantom{0} & $+158.1 \pm 92.7$\phantom{0} & $+154.2 \pm 19.1$ & $8.57$$^G$ & $35.46$ \\
J094750$-$371528$^\odot$ & $+328.8 \pm 2.2$ & $+311.0 \pm 7.2$\phantom{0} & \phantom{0}$-42.2 \pm 148.7$ & \phantom{00}$-2.2 \pm 28.3$ & $12.4$$^G$ & $78.99$ \\
J094808$-$344010$^\dagger$ & $+382.7^{+2.6}_{-2.4}$ & $-327.9 \pm 10.6$ & \phantom{0}$+87.9 \pm 137.8$ & \phantom{0}$+31.9 \pm 23.4$ & $11.4$$^G$ & $48.90$ \\
J111857$+$123442$^\dagger$ & $+79.4^{+2.6}_{-2.8}$ & $-465.4 \pm 5.7$\phantom{0} & $+10.0 \pm 27.1$ & $+11.8 \pm 4.0$ & $1.86$$^E$ & $0.26$ \\
J154936$+$183500$^?$ & $-119.3^{+7.5}_{-7.4}$ & $-426.8 \pm 14.6$ & $+22.8 \pm 16.6$ & $+21.3 \pm 6.9$ & $2.72$$^E$ & $0.89$ \\
J162706$-$091705$^{\star\star}$ & $-327.8 \pm 0.7$ & $-297.2 \pm 12.8$ & $-122.0 \pm 70.7$\phantom{0} & $-154.3 \pm 18.4$ & $12.1$$^G$ & $132.04$ \\
J163927$-$124139$^{\star\star}$ & $-331.4^{+0.3}_{-0.3}$ & $-313.5 \pm 3.6$\phantom{0} & $-205.0 \pm 85.4$\phantom{0} & $-195.0 \pm 21.6$ & $14.3$$^G$ & $92.15$ \\
J170934$-$172853$^\dagger$ & $+106.2^{+1.8}_{-1.9}$ & $-490.0 \pm 12.7$ & $+14.1 \pm 25.7$ & \phantom{00}$-9.7 \pm 26.6$ & $16.5$$^G$ & $4.65$ \\
J190255$+$315942$^\dagger$ & $+142.2^{+1.1}_{-1.1}$ & $-424.3 \pm 2.6$\phantom{0} & $+66.7 \pm 29.4$ & \phantom{0}$+77.2 \pm 24.2$ & $9.91$$^E$ & $3.93$ \\
J220205$+$394913 & $-367.2 \pm 0.4$ & $-349.1 \pm 6.6$\phantom{0} & $-89.0 \pm 95.9$ & $-145.0 \pm 12.5$ & $13.8$$^E$ & $11.57$ \\
J220927$+$415834 & $-338.1 \pm 0.2$ & $-336.0 \pm 5.4$\phantom{0} & $-156.1 \pm 132.6$ & $-193.9 \pm 31.9$ & $15.7$$^E$ & $5.75$ \\
J224412$+$405715$^\dagger$ & $-320.4 \pm 0.6$ & $+345.8 \pm 14.4$ & $-116.7 \pm 97.8$\phantom{0} & $-182.8 \pm 24.3$ & $10.8$$^E$ & $26.40$ \\
J224549$+$394122$^{\odot\dagger}$ & $-278.6^{+0.8}_{-1.0}$ & $+373.3 \pm 6.4$\phantom{0} & $-149.0 \pm 93.4$\phantom{0} & $-259.2 \pm 7.7$\phantom{0} & $8.56$$^E$ & $14.91$ \\
J234033$+$133300$^\times$ & --- & $+56.7 \pm 6.3$ & \phantom{0}$-8.2 \pm 18.1$ & \phantom{0}$+4.7 \pm 6.8$ & $5.07$$^E$ & $1.31$ \\
J235728$+$230226$^\dagger$ & $+42.3 \pm 6.1$ & $-556.7 \pm 13.8$ & $-55.1 \pm 19.7$ & \phantom{0}$-55.3 \pm 11.2$ & $4.23$$^E$ & $0.85$ \\
\hline
\multicolumn{7}{l}{$^a$Median RM values of \citetalias{taylor09} sources within $3^\circ$ radius, with the uncertainties being the standard deviations of those} \\
\multicolumn{7}{l}{\phantom{$^a$}neighbouring sources} \\
\multicolumn{7}{l}{$^b$Galactic contribution to RM from \cite{oppermann15}} \\
\multicolumn{7}{l}{$^c$Neutral hydrogen column density from H~{\sc i} observations} \\
\multicolumn{7}{l}{$^d$Velocity-integrated H~$\alpha$ intensity from the WHAMSS \citep{haffner03,haffner10}} \\
\multicolumn{7}{l}{$^\dagger$Suffers $n\pi$-ambiguity in the \citetalias{taylor09} catalogue} \\
\multicolumn{7}{l}{$^\times$ Unpolarised sources} \\
\multicolumn{7}{l}{$^?$Special case compared to \citetalias{taylor09} catalogue (see Section~\ref{sec:j1549})} \\
\multicolumn{7}{l}{$^{\star\star}$Double point sources} \\
\multicolumn{7}{l}{$^\odot$Extended sources} \\
\multicolumn{7}{l}{$^E$From the Effelsberg-Bonn H~{\sc i} Survey \citep[EBHIS;][]{winkel16}} \\
\multicolumn{7}{l}{$^G$From the Galactic All-Sky Survey \citep[GASS;][]{mcclure-griffiths09,kalberla15}}
\end{tabular}
\end{table*}

\subsection{The Origin of Large Faraday Depths}
Out of our 21 polarised target sources, we found (from RM-Synthesis) that 15 of them have $|\overline\phi| > 200\,{\rm rad\,m}^{-2}$. Such high $|\overline\phi|$ values are peculiar for sources away from the Galactic plane, as is the case for our targets ($|b| > 10^\circ$). While the FD could originate from within the EGSs themselves or from their immediate ambient media, it is challenging to directly confirm this scenario with the available information. We therefore explore the possibility of explaining the FD values from Galactic contributions and/or from foreground galaxy clusters. For the former, as Galactic FD (or RM) structures are often associated with warm and/or cold phases of the interstellar medium \citep[e.g.][]{heiles12}, we looked into their respective tracers (H~$\alpha$ and H~{\sc i}) as an attempt to unveil the origin of the FDs. 

\subsubsection{Comparison with H~$\alpha$ Maps}
Upon inspection of the Wisconsin H-Alpha Mapper Sky Survey \citep[WHAMSS;][]{haffner03,haffner10} images, we found that nine of our high $|\overline\phi|$ EGSs (J083930$-$240723, J084701$-$233701, J090015$-$281758, J091145$-$301305, J092410$-$290606, J093349$-$302700, J093544$-$322845, J094750$-$371528, and J094808$-$344010) lie behind the northern arc of the Gum Nebula, two (J162706$-$091705 and J163927$-$124139) lie behind Sh~2-27, two (J224412$+$405715 and J224549$+$394122) situated close to Sh~2-126, and one (J220205$+$394913) lies behind an H~{\sc ii} filament. All of the above sources are positioned on lines of sight with high velocity-integrated H~$\alpha$ intensities ($I_{{\rm H}\alpha} > 10\,{\rm Rayleighs}$; see column 7 of Table~\ref{table:fg}). Thus, the high $|\overline\phi|$ values of these 14 sources could be attributed to foreground Galactic H~{\sc ii} structures. The only remaining EGS (J220927$+$415834) does not appear to be situated behind prominent H~{\sc ii} structures, with a foreground velocity-integrated H~$\alpha$ intensity of only $5.75\,{\rm R}$. The high FD of this source may have originated from a foreground galaxy cluster (Section~\ref{sec:cluster}).

To assess the link between the H~{\sc ii} structures and the high $|\overline\phi|$, we estimate the regular magnetic field strengths ($B_{\rm reg}$) in those H~{\sc ii} clouds needed to produce the observed $\overline\phi$. We omit the Gum Nebula and Sh 2-27 here, as their magnetic field structures are already studied in detail in \cite{purcell15} and \cite{harvey-smith11} respectively using many of the above-mentioned EGSs. Following \cite{harvey-smith11}, emission measure (EM) and $B_{\rm reg}$ are given by
\begin{gather}
{\rm EM} = 2.75\left(\frac{T_e}{10^4\,{\rm K}}\right)^{0.9}\left(\frac{I_{{\rm H}\alpha}}{\rm R}\right)e^\tau\,{\rm cm}^{-6}\,{\rm pc,}\\
B_{\rm reg} \sim \sqrt{3} B_{{\rm reg, }\parallel} = \sqrt{3}\frac{\phi}{0.81\sqrt{\rm EM}\sqrt{fL}}\,\mu{\rm G,}
\end{gather}
where $T_e$ is the electron temperature, $\tau$ is the optical depth due to dust extinction, $B_{{\rm reg, }\parallel}$ is the strength of the regular magnetic field component parallel to the line of sight, $f$ is the filling factor, and $L$ is the integration path length through the H~{\sc ii} filament (in pc). The relationship between $B_{\rm reg}$ and $B_{{\rm reg},\parallel}$ stem from statistical argument, and it is implicitly assumed that $n_e$ is homogeneous. Furthermore, $B_{{\rm reg, }\parallel}$ is assumed to be uniform in both strength and direction along the lines of sight. In particular, since EM is proportional to <$n_e^2$> while FD is only proportional to <$n_e$>, clumps of free electrons can result in large EM values but only moderate FD values. We also assumed here that the observed $\overline\phi$ of our EGSs come entirely from the H~{\sc ii} structures (i.e.\ zero intrinsic FD contributions). To get an upper limit of $B_{\rm reg}$, we assume a low $T_e$ of $7000\,{\rm K}$ \citep[e.g.][]{peimbert17}, a typical $f = 0.1$ \citep[e.g.][]{harvey-smith11}, and adopted a range of optical depth ($\tau = 0.5$--$1.5$) to obtain
\begin{equation}
B_{\rm reg} \lesssim 2.26\text{--}3.73 \left(\frac{\phi}{{\rm rad\,m}^{-2}}\right) \left(\frac{L}{\rm pc}\right)^{-\frac{1}{2}} \left(\frac{I_{{\rm H}\alpha}}{\rm R}\right)^{-\frac{1}{2}}\,\mu{\rm G.} \label{eq:breg_upper}
\end{equation}
Here, a lower optical depth would lead to a larger coefficient in the above Equation. The only undetermined variable, $L$, can be approximated by simple modelling of the geometry of the individual H~{\sc ii} filament. Sh~2-126 is located at a distance of about $370$--$600\,{\rm pc}$ from us \citep{chen08}, and has an intriguing morphology consisting of filamentary/sheet-like structures with widths of about $50^\prime$, translating to $\sim5$--$9\,{\rm pc}$. Considering that both J224412$+$405715 and J224549$+$394122 lie at the outskirt of this H~{\sc ii} structure, we adopt half of the lower limit in width (i.e.\ $2.5\,{\rm pc}$) as the path length through Sh~2-126. The H~{\sc ii} filament shrouding J220205$+$394913 has an angular width of about $20^\prime$. It has not been studied in detail and thus has an unknown distance, but its spatial proximity and similarity in radial velocity ($v_{\rm LSR} \approx -9.0\,{\rm km\,s}^{-1}$ from WHAMSS) with nearby H~{\sc ii} structures Sh~2-118 and Sh~2-123 suggest physical associations among these objects. The latter two clouds have kinematic distances of $3.8\,{\rm kpc}$ \citep{russeil03}, which if taken as the distance to the H~{\sc ii} filament would yield a physical width of $22\,{\rm pc}$. We adopt this as the path length through this filament. Note that $B_{\rm reg}$ is only weakly sensitive to $L$, as an over-/under-estimation of the latter by 10 times would only result in the former being weaker/stronger by a factor of 3.2, which would not affect our order-of-magnitude estimation here.

Substituting in the adopted values of $L$ for each H~{\sc ii} structure, as well as $I_{{\rm H}\alpha}$ from the WHAMSS (see Table~\ref{table:fg}) and $\phi$ as our $\overline\phi$ values into Equation~\ref{eq:breg_upper}, we obtain $B_{\rm reg} \lesssim 89$--$171\,\mu{\rm G}$ for Sh~2-126 and $52$--$86\,\mu{\rm G}$ for the H~{\sc ii} filament in front of J220205$+$394913. The field strengths here are an order of magnitude higher than that in typical Galactic H~{\sc ii} regions \citep[$\sim 1$--$36\,\mu{\rm G}$; e.g.][]{heiles81, gaensler01, harvey-smith11, rodriguez12}, though note that these H~{\sc ii} filaments might not be typical H~{\sc ii} regions. Since our crude assumptions above would yield upper limits in field strengths, and for these two clouds we only have very rough estimates on the physical scales, we cannot draw a concrete conclusion on whether these two H~{\sc ii} structures can contribute to the bulk of the observed $|{\overline\phi}|$ of the three target sources.

\subsubsection{Comparison with H~{\sc i} Column Densities}

We also looked into the Galactic H~{\sc i} column densities ($N_{\mathrm{H}\textsc{i}}$) towards our target sources, using the result from the Effelsberg-Bonn H~{\sc i} Survey \citep[EBHIS;][]{winkel16} for the northern sky and the Galactic All-Sky Survey \citep[GASS;][]{mcclure-griffiths09,kalberla15} for the southern hemisphere. The foreground $N_{\mathrm{H}\textsc{i}}$ values for our target sources are listed in column 6 of Table~\ref{table:fg}. However, we do not see any clear trends between $|\overline\phi|$ and $N_{\mathrm{H}\textsc{i}}$.

\subsubsection{Foreground Galaxy Clusters \label{sec:cluster}}

We explore the possibility of high FD stemming from the hot magnetised intracluster medium of foreground galaxy clusters \citep[see][]{govoni04}, which can have $|{\rm FD}|$ contributions to embedded / background polarised sources of $\sim 100\,{\rm rad\,m}^{-2}$ \citep[e.g.][]{taylor01,bonafede09,govoni10}. The NASA/IPAC Extragalactic Database (NED) was consulted for galaxy clusters within $2^\circ$ of our 15 high $|\overline\phi|$ target sources, and we found matches for the six sources below.

\begin{enumerate}
\item J084701$-$233701 at $z = 0.0607 \pm 0.0001$ \citep{2mass} with $\overline\phi = +384.8 \pm 2.0\,{\rm rad\,m}^{-2}$ is situated at $34\farcm3$ away from Abell S0613 and $96\farcm5$ away from PSZ1 G246.45$+$13.16. The former galaxy cluster is background to our target \citep[$z = 0.0740$;][]{chow14}, and therefore cannot contribute to the high FD. The latter is a Sunyaev-Zel'dovich cluster candidate \citep{planck14} with poorly constrained parameters.
\item J093349$-$302700 ($z$ unknown) with $\overline\phi = +341.6 \pm 0.8\,{\rm rad\,m}^{-2}$ is accompanied by Abell 3421 at $92\farcm4$ away and Abell S0618 at $92\farcm9$ away. Both of the clusters do not have constrained $z$ nor angular sizes.
\item J094808$-$344010 ($z$ unknown) with $\overline\phi = +382.7^{+2.5}_{-2.4}\,{\rm rad\,m}^{-2}$ is $81\farcm1$ away from Abell 3428. This cluster has a photometric redshift of $z = 0.0601$ \citep{coziol09}, but without any reported angular sizes.
\item J220205$+$394913 ($z$ unknown) with $\overline\phi = -367.2 \pm 0.4\,{\rm rad\,m}^{-2}$ is situated next to ZwCl 2200.7$+$3752 at $103\farcm1$ away. This cluster has a diameter of only $58^\prime$ \citep{zwicky61}, and therefore cannot contribute to the FD of our target.
\item J220927$+$415834 at $z = 0.512 \pm 0.029$ \citep[photometric;][]{sdssdr14} with $\overline\phi = -338.1 \pm 0.2\,{\rm rad\,m}^{-2}$ is neighbouring UGCL 467 (also known as ZwCl 2207.8$+$4114) at $30\farcm2$ away. This foreground cluster ($z = 0.0166$; $1^\prime = 21\,{\rm kpc}$) has a diameter of $164^\prime$ \citep{baiesi84}, and could be the prime contributor to FD of J220927$+$415834 given that we could not identify any clear foreground Galactic structures from H $\alpha$ nor H~{\sc i} above.
\item J224412$+$405715 at $z = 1.171$ \citep{fermiagn2} with $\overline\phi = -320.4 \pm 0.6\,{\rm rad\,m}^{-2}$ is accompanied by two galaxy clusters -- 1RXS J223758.3$+$410109 ($70\farcm8$ away) and Sunyaev-Zel'dovich cluster candidate PSZ1 G097.52$-$14.92 ($77\farcm1$ away). Neither of them have reported $z$ nor cluster diameter.
\end{enumerate}

To summarise, J220927$+$415834 (for which we could not find any foreground Galactic H $\alpha$ or H {\sc i} structures to) may have attained its high FD from the foreground galaxy cluster UGCL 467. We cannot confidently attribute the high FD of the rest of our target sources to foreground clusters, given the ill-constrained parameters, particularly redshifts,  to the sources themselves and/or to the foreground clusters.

\subsection{The Nature of Faraday Complexity}
\subsubsection{Definition of Faraday Complex Sources \label{sec:complex_def}}
We find it necessary to formally define Faraday complex sources before proceeding further. The main reason is to facilitate comparisons with the literature, as a growing number of broadband spectro-polarimetric studies of EGSs choose to extract the flux densities of their samples by integrating within a source region \citep[e.g.,][]{anderson16,osullivan17}. While this would be similar to the strategies we adopted for our point sources and extended sources, it is in contrast to our spatial doubles, for which we fitted two Gaussian functions to each image (per frequency channel and per Stokes parameter; Section~\ref{sec:fullbandreduction}) and analysed the two spatial components independently. In other words, although we may be able to identify small differences in FD between two spatial components, the same source may be classified as Faraday simple when the spatial information is discarded. We therefore carefully define Faraday complexity for our target sources here to match the expected outcome if our sources were not spatially resolved. Also, in addition to angular resolution (see Appendix~\ref{sec:spatial}), we note that whether the Faraday complexity of a source can be correctly identified can also depend on the $S/N$ ratio \citep[e.g.][]{anderson15,osullivan17} and $\lambda^2$ coverage \citep[e.g.][]{anderson16}.

We therefore define Faraday complex sources as follows. From RM-Synthesis, an unresolved or extended source is considered as Faraday complex if it is decomposed into multiple Faraday components, or the only Faraday component is Faraday thick. Here, we define ``Faraday thick component'' as one with the fitted FWHM ($\delta\phi$) at least 10 per cent more than the theoretical FWHM of the RMTF ($\delta\phi_0$), while ``Faraday thin component'' is one with $\delta\phi$ less than $1.1$ times of $\delta\phi_0$. For a spatial double source, it is deemed Faraday complex if at least one of the spatial components is further divided into multiple Faraday components, or one/each of the spatial components hosts a Faraday thick component, or each spatial component contains one and only one Faraday thin component but the FDs of these two components are separated by more than 37 per cent of $\delta\phi_0$ (the choice of this factor is explained below). On the other hand, from \textit{QU}-fitting a spatially unresolved or extended source is defined as Faraday complex if its best-fit model is not single thin (1T), while a double source is categorised as Faraday complex if either/both of the spatial components is/are best-fitted by models other than single thin, or both spatial components are best characterised by the single thin model but the difference in FDs of the two Faraday simple components is larger than 37 per cent of the $\delta\phi_0$ from RM-Synthesis (again, the choice of this factor is explained below). 

As mentioned above, the most critical part of this formal definition here is for spatial double sources, particularly for cases where each spatial component hosts a Faraday thin component. In such cases, the two spatially resolved Faraday components could be indistinguishable from a single Faraday thin component if we discard the spatial information by combining them within a source integration region. While previous works showed by simulations that Faraday components with FDs separated by less than $\approx 50$--$100$ per cent of $\delta\phi_0$ cannot be confidently distinguished by RM-Synthesis and \textit{QU}-fitting \citep[e.g.][]{farnsworth11,sun15,schnitzeler18,miyashita17}, we chose to adopt a smaller cutoff value of 37 per cent here. This is because although the two Faraday thin components cannot be separated if they are situated too close together in Faraday space, the two combined could be identified as a single Faraday thick component. We calculated for the simplest case of adding two Faraday thin components with equal amplitudes together, and found that when they are separated by about 37 per cent of $\delta\phi_0$ the combined function resembles a Gaussian function with $\delta\phi$ being 1.1 times of $\delta\phi_0$, satisfying our definition of Faraday thick component above. Nonetheless, the choice of this cutoff value would not affect the results of our work here, as the most extreme case we have is J093544$-$322845 in RM-Synthesis, with the two Faraday thin components separated by 8.7 per cent of $\delta\phi_0$ only.

\subsubsection{The Physical Origin of Faraday Complexity}

One of the major strengths of radio broadband spectro-polarimetric observations is its ability to decompose spatially unresolved sources (e.g.\ EGSs) into multiple Faraday components. These components could be located anywhere in the volume traced by the telescope beam, both parallel or perpendicular to the line of sight. This opens up the possibility of identification or even study of discrete physical regions that are spatially unresolved by the observations, but this would require prior studies associating the Faraday components with spatial components for a sample of spatially resolved sources. There appears to be some correspondences between the number of Faraday and spatial components of EGSs \citep[e.g.][both with angular resolution of $\sim 1^{\prime\prime}$]{anderson16,osullivan17}. This motivates us to carry out similar investigations to our sample of EGSs below.

We first look at sources that are spatially resolved with our $\sim 45^{\prime\prime}$ beam. The only such sources that are resolved into multiple Faraday components are J224549$+$394122 from RM-Synthesis, and J092410$-$290606 and J162706$-$091705 from \textit{QU}-fitting. All three of them host two Faraday components each, with J224549$+$394122 being resolved into FR II morphology and the remaining two as double unresolved components. Interestingly, the two Faraday components for each of the spatial doubles originate from just one of the two spatial components (J092410$-$290606a and J162706$-$091705b respectively). J092410$-$290606b is polarised, but its Faraday component is indistinguishable from one of the two from J092410$-$290606a. On the other hand, J162706$-$091705a is not polarised (below $6\sigma$ limit of $3$ per cent). For the remaining three spatial doubles (J091145$-$301305, J093544$-$322845, and J163927$-$124139), the sources are not resolved into multiple Faraday components according to our definition above in Section~\ref{sec:complex_def}. However, since we analysed the spatial components individually in both RM-Synthesis and \textit{QU}-fitting, we can still obtain the difference in FD between the two components, and compute what $\lambda^2$ coverages are required to resolve them into two Faraday components if these sources were spatially unresolved. From RM-Synthesis and \textit{QU}-fitting, our spatial doubles have differences in FD of $2.4$--$7.6\,{\rm rad\,m}^{-2}$ and $2.7$--$15.5\,{\rm rad\,m}^{-2}$, respectively. Assuming that in both analyses we can distinguish Faraday components separated by more than 50 per cent of the theoretical $\delta\phi_0$ in RM-Synthesis \citep[e.g.][]{schnitzeler18}, a $\lambda^2$ coverage of more than $0.11$--$0.72\,{\rm m}^2$ would be needed to resolve our spatial doubles into the multiple Faraday components. These translate to frequency coverages from $1\,{\rm GHz}$ down to $660$ and $330\,{\rm MHz}$, respectively. From this, we argue that for spectro-polarimetric studies of EGSs in GHz regime, we should \emph{not} combine multiple spatial components together with flux integration regions. This is because the spatially resolved Faraday components would then become a single unresolved Faraday component, leading to loss of physical information of the sources. We draw similar conclusions in Appendix~\ref{sec:spatial} for our spatially extended sources, where we found that the \textit{QU}-fitting results after spatial flux integrations differs from our spatially resolved RM-Synthesis analysis.

Furthermore, we searched for Faint Images of the Radio Sky at Twenty-Centimeters \citep[FIRST; angular resolution $\approx 5^{\prime\prime}$;][]{becker95}, as well as Very Long Baseline Interferometry \citep[VLBI; angular resolution $\sim {\rm mas}$;][]{fey97,fey00} total intensity images of all of our target sources. We found that four of them have existing higher angular resolution radio images. These sources are discussed individually below.
\begin{enumerate}
\item J111857$+$123442 (4C +12.39) is composed of two Faraday thin components in both of our analysis. In the FIRST image there is a hint of a fainter spatial component $10^{\prime\prime}$ to the northwest of the main component. At VLBI resolution the source is extended at $2.3\,{\rm GHz}$, and is resolved into two spatial components at $8.6\,{\rm GHz}$.
\item J154936$+$183500 (4C +18.45; the special case; Section~\ref{sec:j1549}) consists of three Faraday components in both RM-Synthesis and \textit{QU}-fitting. These Faraday components have vastly different FD values ($-315.3 \pm 2.1$, $-31.8 \pm 3.5$, and $+81.3 \pm 2.8\,{\rm rad\,m}^{-2}$ from RM-Synthesis; similar to that from \textit{QU}-fitting). This source is also spatially resolved into three components in the FIRST image -- two bright blobs together resembling an FR II radio galaxy with an angular scale of about $15^{\prime\prime}$ \citep[corresponding to a projected physical scale of about 130\,kpc at $z = 1.442$;][]{hewitt87}, and a third faint point source situated about $40^{\prime\prime}$ away to the southwest.
\item J170934$-$172853 is represented by two Faraday components in our RM-Synthesis and \textit{QU}-fitting analysis. The source appears in the VLBI image at $2.3\,{\rm GHz}$ as two spatial components, with the brighter one with flux density of about $500\,{\rm mJy}$ and the dimmer one situated about $10\,{\rm mas}$ away to the southeast with flux density of about $10\,{\rm mJy}$. The brighter component can be further resolved into two components at $8.6\,{\rm GHz}$, with component 1 at about $300\,{\rm mJy}$ and to the southeast by $3\,{\rm mas}$ component 2 at about $5\,{\rm mJy}$.
\item J190255$+$315942 (3C 395) is found to have two Faraday thin components in our analysis. In the $2.3\,{\rm GHz}$ VLBI image it is consisted of two spatial components separated by about 15\,mas \citep[projected distance of about 100\,pc at $z = 0.635$;][]{hewitt87}, with a faint structure connecting the two. The two components can also be seen in the $8.6\,{\rm GHz}$ VLBI image.
\end{enumerate}

From above, there appears to be a good association between the number of Faraday components identified from our 1--2\,GHz observations and the number of spatial components resolved at $5^{\prime\prime}$ (FIRST) or mas (VLBI) resolutions. A caveat here is that because of the missing short \textit{uv}-spacing, there could be missing flux from structures on large angular scales, particularly in the VLBI images. Note that this suggested association between the number of spatial and Faraday components is only speculative, and requires confirmation from high angular resolution spectro-polarimetric studies. Indeed, a more comprehensive study on the connection between Faraday components and structures of EGSs at different angular and physical scales, as well as for different source types, would be necessary before we can confidently interpret their Faraday complexities from low angular resolution observations alone.

Finally, it has been suggested that lines of sight with Galactic H~{\sc i} column density of $1.4$--$1.65 \times 10^{20}\,{\rm cm}^{-2}$ may pass through magnetised plasma in the Milky Way which could cause observed Faraday complexities in background EGSs \citep{anderson15}. This would imply that the turbulence scale of the magneto-ionic medium causing such complexities is less than $\sim 5\,{\rm pc}$ assuming a distance to the far side of the Milky Way of $23.5\,{\rm kpc}$ with their angular resolution of $\sim 45^{\prime\prime}$. All of our target sources have foreground H~{\sc i} column densities higher than the above-mentioned range (Table~\ref{table:fg}), with J111857$+$123442 having the lowest of $1.86 \times 10^{20}\,{\rm cm}^{-2}$. It is represented by double Faraday thin components in both RM-Synthesis and \textit{QU}-fitting, with differences in FD of about $100\,{\rm rad\,m}^{-2}$. This source is resolved into two spatial components separated by about $10^{\prime\prime}$ in FIRST (see above). At such a small angular scale, the Milky Way contribution to FD is not expected to vary by such a large amount. We suggest that for this source, Faraday complexity is not caused by the magneto-ionic medium in the Milky Way.

\subsubsection{Faraday Complexity Statistics}
Our RM-Synthesis and \textit{QU}-fitting results show respectively that eight (38 per cent) and 14 (67 per cent) out of the 21 polarised target sources are Faraday complex. We briefly discuss the difference between these two algorithms in Appendix~\ref{sec:rm_qu_compare}. The RM-Synthesis fraction is similar to the 29 per cent (12 out of 42) obtained from the RM-Synthesis analysis on ATA data of bright radio sources in $1$--$2\,{\rm GHz}$ \citep[angular resolution $\sim 100^{\prime\prime}$;][]{law11}. This similarity may be because of the similar $\lambda^2$ coverages, as well as the high signal-to-noise ratio in polarisation, in both studies. In contrast, \cite{anderson15} reported with their $1.3$--$2.0\,{\rm GHz}$ study at an angular resolution of $\sim 1^\prime$ that only 12 per cent (19 out of 160) of their polarised sources appeared to be Faraday complex with their observational setup. This can be attributed to the lower signal-to-noise ratio in PI ($\lesssim 10$) of some of their target sources. As they suggested in their paper, sources that are genuinely Faraday complex might appear Faraday simple in the low S/N regime.

There are spectro-polarimetric studies of EGSs at other wavelengths that reported a much higher fraction of Faraday complex sources. \cite{pasetto18} found by \textit{QU}-fitting analysis that, all of their 14 high RM sources are Faraday complex with their $4$--$12\,{\rm GHz}$ observations (angular resolution $\lesssim 1^{\prime\prime}$), though this could be biased due to their source selection criteria. They chose sources that are unpolarised in the NVSS at $1.4\,{\rm GHz}$ but polarised at higher frequencies, which could be due to bandwidth depolarisation in the NVSS at specific $|{\rm RM}|$ ranges ($\approx 350$ or $\gtrsim 1000\,{\rm rad\,m}^{-2}$) and/or Faraday depolarisation due to complexities. Nonetheless, \cite{osullivan17} reported that 90 per cent (90 out of 100) of their targets are Faraday complex from their $1$--$3\,{\rm GHz}$ observations with angular resolution of $\sim 10^{\prime\prime}$, also with \textit{QU}-fitting analysis. This is similar to the findings of \cite{anderson16}, who observed at $1.3$--$10\,{\rm GHz}$ (with angular resolution of $\sim 1$--$10^{\prime\prime}$) a total of 36 EGSs selected such that, based on archival narrowband $1.4\,{\rm GHz}$ data, half of the sample are Faraday simple and the other half are Faraday complex. Their broadband studies with RM-Synthesis concluded that 97 per cent (35 out of the 36) of their sample turns out to be Faraday complex in the observed $\lambda^2$ range, with the remaining one consistent with being unpolarised. By re-analysing their data at different $\lambda^2$ coverages, they suggested that the detection of Faraday complexity of EGSs could be hindered by limited $\lambda^2$ ranges. This suggests that many of our Faraday simple sources could become Faraday complex if they are observed at a wider $\lambda^2$ range with sufficient $S/N$ ratio.

\section{Conclusion} \label{sec:conclusion}
With new broadband spectro-polarimetric observations of 23 $n\pi$-ambiguity candidates with the VLA in L-band, we revealed nine out-\textit{liars} (sources that suffer from $n\pi$-ambiguity in the NVSS RM catalogue). By comparing the statistics of their observed parameters with that of the 11 outliers (sources with reliable ${\rm RM}_{\rm TSS09}$), we find noticeable differences between the two classes in $\alpha_{\rm L}$, $p_{\rm TSS09}$, $|{\rm RM}_{\rm TSS09} - {\rm RM}_{3^\circ}|$, $\Delta/\sigma = |{\rm RM}_{\rm TSS09} - {\rm RM}_{3^\circ}|/\sigma_{3^\circ}$, and Faraday complexities. In particular, we find $\Delta/\sigma$, which is a measure of how much a source's RM deviates from the RMs of its surrounding sources, to be a good diagnostic for $n\pi$-ambiguity in the NVSS RM catalogue. There is an apparent cutoff at $\Delta/\sigma \approx 2.85$ between the two populations, which we used to estimate that there are at least 50 $n\pi$-ambiguity sources in the \citetalias{taylor09} catalogue out of the total of 37,543 sources. This is an important result for us to gauge the reliability of the \citetalias{taylor09} catalogue, and merits further studies to verify these $n\pi$-ambiguity candidates. We further identified two sources that are polarised in \citetalias{taylor09} at 0.5--0.6 per cent levels, but are unpolarised (below the $6\sigma$ cutoffs of $\approx 0.07$ per cent) in our new broadband observations. These two sources have motivated a detailed study on the effects of the off-axis instrumental polarisation in the NVSS RM catalogue, presented in the companion Paper II.

We found that 15 of our target sources have large $|\overline{\phi}| > 200\,{\rm rad\,m}^{-2}$ despite being situated away from the Galactic plane ($|b| > 10^\circ$). 14 of them are found to be lying behind Galactic H~{\sc ii} structures, which are likely the prime contributors to the observed high $|\overline{\phi}|$ of these sources. The only remaining source, J220927$+$415834, is found to be background to the galaxy cluster UGCL 467, which is the most likely explanation of its high $|\overline{\phi}|$.

Finally, we studied the Faraday complexities of our target sources with our broadband 1--2\,GHz observations. We found good correspondence between the number of identified Faraday components from our analysis with the number of spatial components in total intensities at $\approx 5^{\prime\prime}$ and milli-arcsecond resolutions in FIRST and VLBI images, respectively. However, this speculated associations between the Faraday and spatial components require confirmation from future polarisation studies at high angular resolution. In our sample of 21 polarised sources, eight (38 percent) and 14 (67 per cent) are Faraday complex from our RM-Synthesis and \textit{QU}-fitting analysis respectively. The former value agrees with the 29 per cent reported by \cite{law11} with their RM-Synthesis study of EGSs at similar frequency range. We noted that if our target sources are re-observed with a wider $\lambda^2$ coverage than that of our L-band observations here, many of our current Faraday simple sources will likely become Faraday complex at sufficient signal-to-noise ratio in polarisation.

\section*{Acknowledgements}
This is a pre-copyedited, author-produced PDF of an article accepted for publication in the Monthly Notices of the Royal Astronomical Society following peer review. The version of record is available at: xxxxxxx. We thank the anonymous referee for the comments, especially for the suggestion to separate our original manuscript into two stand-alone publications, which have improved the clarity of the papers. We also thank Rainer Beck for his careful reading and valuable suggestions and comments as the MPIfR internal referee, and Aristeidis Noutsos, Shane O'Sullivan, and Dominic Schnitzeler for insightful discussions about this project. Y.K.M.\ was supported for this research by the International Max Planck Research School (IMPRS) for Astronomy and Astrophysics at the Universities of Bonn and Cologne. Y.K.M.\ acknowledges partial support through the Bonn-Cologne Graduate School of Physics and Astronomy. A.B.\ acknowledges financial support by the German Federal Ministry of Education and Research (BMBF) under grant 05A17PB1 (Verbundprojekt D-MeerKAT). A.S.H.\ and S.K.B.\ acknowledge support by NASA through grant number HST-AR-14297 to Haverford College from Space Telescope Science Institute, which is operated by AURA, Inc.\ under NASA contract NAS 5-26555. A.S.H.\ is partially supported by the Dunlap Institute, which is funded through an endowment established by the David Dunlap family and the University of Toronto. The National Radio Astronomy Observatory is a facility of the National Science Foundation operated under cooperative agreement by Associated Universities, Inc. This research has made use of the NASA/IPAC Extragalactic Database (NED) which is operated by the Jet Propulsion Laboratory, California Institute of Technology, under contract with the National Aeronautics and Space Administration. The Wisconsin H-Alpha Mapper and its Sky Survey have been funded primarily through awards from the U.S.\ National Science Foundation.

\bsp

\bibliography{ms}

\appendix

\section{Spatially Resolved Sources} \label{sec:spatial}

In our new VLA D array observations, two of our target sources are spatially resolved, namely J094750$-$371528 and J224549$+$394122. The former is identified as a galaxy \citep[PGC 626051;][]{paturel89} at $z=0.0411$ \citep{6dFGS}, while the latter is an FR II radio galaxy at $z=0.0811$ and is well studied in the radio regime \citep[commonly known as 3C 452; e.g.][]{black92,harwood17}. For these two sources, we formed another set of Stokes \textit{I}, \textit{Q}, and \textit{U} images for detailed spatial analysis. Channel images were again formed by binning $4\,{\rm MHz}$ of visibilities with identical algorithm and weighting scheme as the full band data (Section~\ref{sec:fullbandreduction}). The only difference is that the images formed here are re-smoothed to a common beam size ($210^{\prime\prime} \times 50^{\prime\prime}$ for J094750$-$371528 and $70^{\prime\prime} \times 55^{\prime\prime}$ for J224549$+$394122) for each source. This step is necessary for the derivation of spectral index and FD maps in the following.

\begin{figure*}
\includegraphics[width=\doublecolumnwidth]{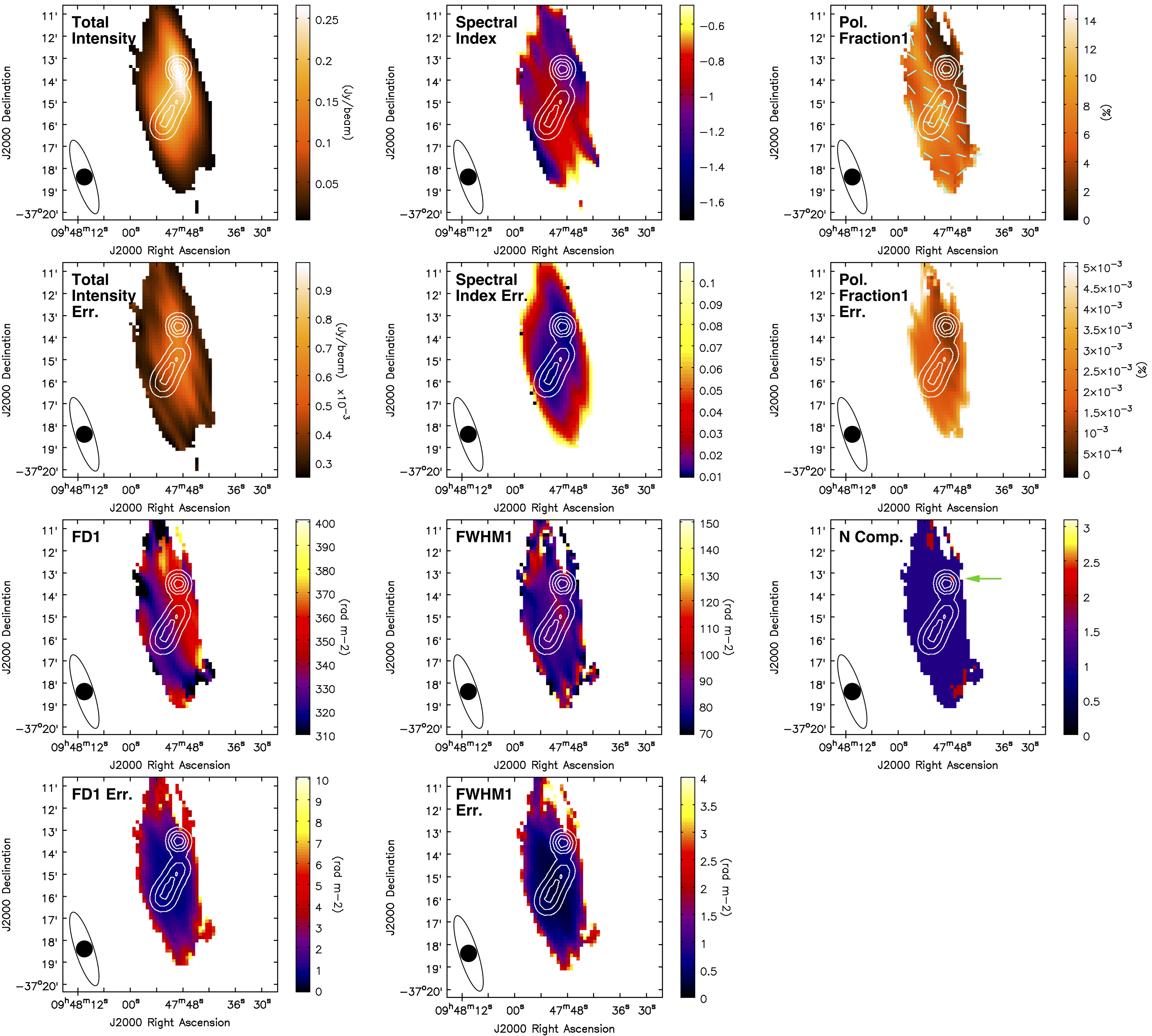}
\caption{Images of J094750$-$371528, with their associated uncertainties. Only pixels where the Stokes \textit{I} values are greater than $6\sigma$ in all of the individual channels are shown. The white contours represent NVSS Stokes \textit{I} map at $[0.2, 0.4, 0.6, 0.8] \times 188.6\,{\rm mJy\,beam}^{-1}$. The beam sizes of the maps are shown in the lower left of each panel, with the open and filled ellipses representing that of our new observation and NVSS, respectively. We also plot the intrinsic polarisation $B$-orientations ($B_0 = E_0 + 90^\circ$; corrected for Faraday rotation) in the polarisation fraction panel as cyan lines. The green arrow in the ``N Comp.'' panel points to the region with a marginal detection of a secondary Faraday component as discussed in text. \label{fig:j0947}}
\end{figure*}

\begin{figure*}
\includegraphics[width=\doublecolumnwidth]{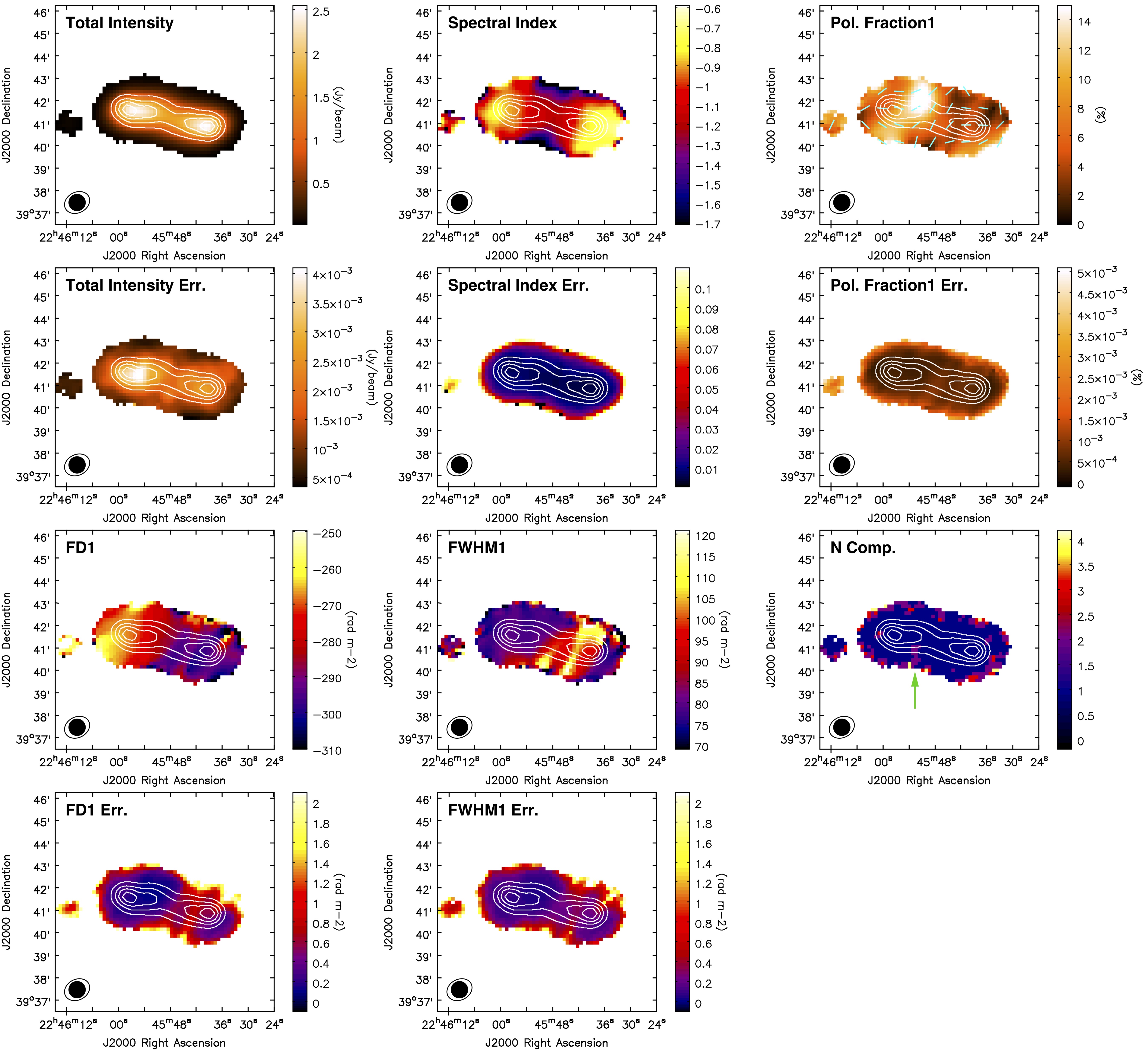}
\caption{Same as Figure~\ref{fig:j0947}, for J224549$+$394122. The white contours here represent NVSS Stokes \textit{I} map at $[0.2, 0.4, 0.6, 0.8] \times 1823.0\,{\rm mJy\,beam}^{-1}$, and the green arrow in the ``N Comp.'' panel points to the finger-like patch with a marginal detection of a secondary Faraday component as discussed in text. \label{fig:j2245}}
\end{figure*}

With the smoothed 4\,MHz channel images of J094750$-$371528 and J224549$+$394122, we generated maps of Stokes \textit{I} total intensity at $1.4\,{\rm GHz}$ ($S_{1.4\,{\rm GHz}}$) and spectral index ($\alpha_{\rm L}$) by fitting simple power law to each individual pixels in the maps:
\begin{equation}
S_\nu = S_{\rm 1.4\,GHz} \cdot \left( \frac{\nu}{1.4\,{\rm GHz}}\right)^{\alpha_{\rm L}}{\rm ,}
\end{equation}
where $\nu$ represents the observed frequency. Only pixels where the Stokes \textit{I} values are larger than $6\sigma$ in all channels are fitted. In addition, we performed RM-Synthesis for each pixel as per Section~\ref{sec:rmsyn}. The maps of number of Faraday components (N Comp.), as well as that of $p$, FD, and FWHM of the strongest Faraday component, were created. All of the above-mentioned maps, along with their uncertainties (if applicable) are presented in Figures~\ref{fig:j0947} (for J094750$-$371528) and \ref{fig:j2245} (for J224549$+$394122). We do not show the maps of the weaker Faraday component(s), since for all cases here the secondary component is weak and can be artefacts in the Faraday spectra instead of real signals. Nonetheless, we report here the most significant secondary components for both sources. For J094750$-$371528, there are two pixels within the northern component of the source where secondary Faraday components is seen, which has $p \approx 0.8$ per cent, $\phi \approx +502\,{\rm rad\,m}^{-2}$, and ${\rm PA}_0 \approx -36^\circ$, while for J224549$+$394122, there is a finger-like patch of secondary component to the south from the eastern jet. This patch has $p \approx 0.2$ per cent, $\phi \approx -416\,{\rm rad\,m}^{-2}$, and ${\rm PA}_0 \approx +80^\circ$.

The polarisation maps we obtained here allow us to make interesting comparisons with our results from the main text, where we used flux integration regions for analysis, which discarded all spatial information of these two sources. We note that J094750$-$371528 is mostly Faraday thin over the entire spatial extent (i.e.\ ${\rm FWHM}_1 \approx {\rm FWHM}_0 = 77\,{\rm rad\,m}^{-2}$). There is also a significant FD gradient from southeast ($\approx +320\,{\rm rad\,m}^{-2}$) to northwest ($\approx +350\,{\rm rad\,m}^{-2}$). This source was found to be Faraday thick in both our RM-Synthesis and \textit{QU}-fitting results. The analysis here shows that the Faraday thickness is caused by spatial variations of FD across the sky plane (i.e.\ side-to-side variations in FD), but not due to Faraday rotation within the synchrotron-emitting medium (i.e.\ back-to-front changes in FD). The polarisation maps of J224549$+$394122 show much richer structures. First, we notice a bright blob in polarisation map where $p \approx 17$ per cent, which coincides spatially with a knot-like structure in the radio jet seen from \cite{harwood17}. Second, there is an FD gradient from east ($\approx -270\,{\rm rad\,m}^{-2}$) to west ($\approx -300\,{\rm rad\,m}^{-2}$), with an FD filament with $\phi \approx -290\,{\rm rad\,m}^{-2}$ in the western part of the source. On this filament we also find a lower $p \approx 4$ per cent and wider ${\rm FWHM}_1 \approx 100\,{\rm rad\,m}^{-2}$ than the surroundings. Similar structures in RM maps were also seen in other systems (e.g. 3C 270, 3C 353, 4C 35.03, and M 84), with those ``RM-bands'' having differences in RM from the off-band regions of $\approx 10$--$50\,{\rm rad\,m}^{-2}$ \citep{guidetti11}.

\section{Comparison between RM-Synthesis and \textit{QU}-fitting Results} \label{sec:rm_qu_compare}

\begin{table}
\centering
\caption{Comparison between RM-Synthesis and \textit{QU}-fitting on the Determination of Faraday Complexities \label{table:rmqu}}
\begin{tabular}{lcc}
\hline
\multicolumn{1}{c}{Source} & RM-Synthesis & \textit{QU}-fitting \\
\multicolumn{1}{c}{(NVSS)} & Results & Results \\
\hline
\multicolumn{3}{c}{\textbf{Sources with Differing Results}}\\
\hline
J022915$+$085125 & Thick & Double \\
J092410$-$290606a & Thin & Double \\
J092410$-$290606b & Thin & Thick \\
J093349$-$302700 & Thin & Thick \\
J162706$-$091705b & Thin & Double \\
J163927$-$124139b & Thin & Thick \\
J170934$-$172853 & Double & Double Thick \\
J224412$+$405715 & Thin & Thick \\
J224549$+$394122 & Double Thick & Thick \\
J235728$+$230226 & Thin & Double \\
\hline
\multicolumn{3}{c}{\textbf{Sources with Agreeing Results}}\\
\hline
J083930$-$240723 & \multicolumn{2}{c|}{Thin} \\
J084701$-$233701 & \multicolumn{2}{c|}{Thin} \\
J090015$-$281758 & \multicolumn{2}{c}{Thin} \\
J091145$-$301305a & \multicolumn{2}{c|}{Thin} \\
J091145$-$301305b & \multicolumn{2}{c|}{Thin} \\
J093544$-$322845a & \multicolumn{2}{c}{Thin} \\
J093544$-$322845b & \multicolumn{2}{c|}{Thin} \\
J094750$-$371528 & \multicolumn{2}{c|}{Thick} \\
J094808$-$344010 & \multicolumn{2}{c}{Double} \\
J111857$+$123442 & \multicolumn{2}{c|}{Double} \\
J154936$+$183500 & \multicolumn{2}{c|}{Triple} \\
J163927$-$124139a & \multicolumn{2}{c}{Thin} \\
J190255$+$315942 & \multicolumn{2}{c|}{Double} \\
J220205$+$394913 & \multicolumn{2}{c|}{Thin} \\
J220927$+$415834 & \multicolumn{2}{c}{Thin} \\
\hline
\end{tabular}
\end{table}

We have presented in Sections~\ref{sec:rmsyn} and \ref{sec:qu} the results from RM-Synthesis and \textit{QU}-fitting, respectively, and noted the discrepant results for several of our target sources from the two analysis methods. This allows comparisons of the two algorithms for the study of polarised EGSs in the frequency range of 1--2\,GHz.

We have divided our RM-Synthesis and \textit{QU}-fitting results into five classes -- thin, double, triple, thick, and double thick. The former three corresponds to one, two, and three unresolved Faraday components in RM-Synthesis, and 1T, 2T, and 3T models in \textit{QU}-fitting, respectively, while the latter two (thick and double thick) maps to single and double resolved Faraday components in RM-Synthesis, and 1B, 1B+fg, 1Ed, 1Id, or 1Id+fg for thick, and 2B or 2B+fg for double thick in \textit{QU}-fitting, respectively. We then compared if the two algorithms agreed on the source class, with the results listed in Table~\ref{table:rmqu}. Out of the 25 polarised sources in our sample (spatial doubles are counted as two distinct sources), 15 have agreeing results, while 10 are categorised into different classes. Further studies of these sources at a wider $\lambda^2$ coverage are needed to determine whether RM-Synthesis or \textit{QU}-fitting are more reliable in uncovering the Faraday complexities of these sources correctly (Ma et al.\ in prep). Nonetheless, we note that when RM-Synthesis identifies a source as Faraday complex, it is definitely so in \textit{QU}-fitting, but the converse is not true.

\section{Online Supplementary Materials} \label{sec:online}

We include here plots of the \textit{QU}-fitting results in Figure~\ref{fig:qufit}. Plots of Stokes $q = Q/I$ and $u = U/I$, along with polarisation fraction ($p$) and polarisation position angle (PA) are shown.

\begin{figure*}
\includegraphics[width=\doublecolumnwidth]{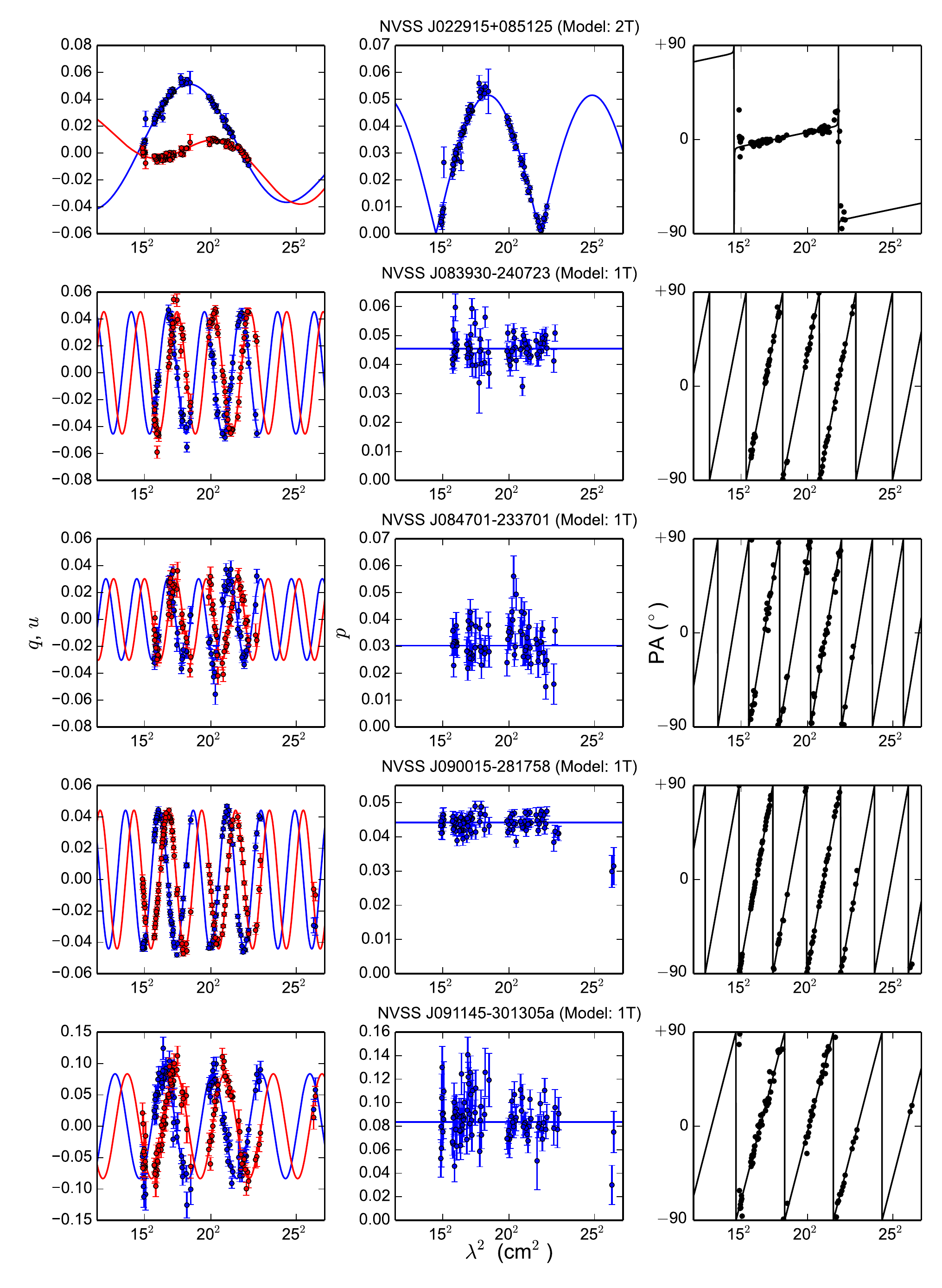}
\caption{\textit{QU}-fitting results of our polarised target sources, showing the best-fit model from our analysis. Each source spans one of the rows in the figure. The Stokes $q = Q/I$ and $u = U/I$ values are plotted in the left column in blue and red respectively, with polarisation fraction ($p$) shown in the middle column, and PA in the right column. Note that the error bars in PA are not shown here.\label{fig:qufit}}
\end{figure*}

\begin{figure*}
\ContinuedFloat
\includegraphics[width=\doublecolumnwidth]{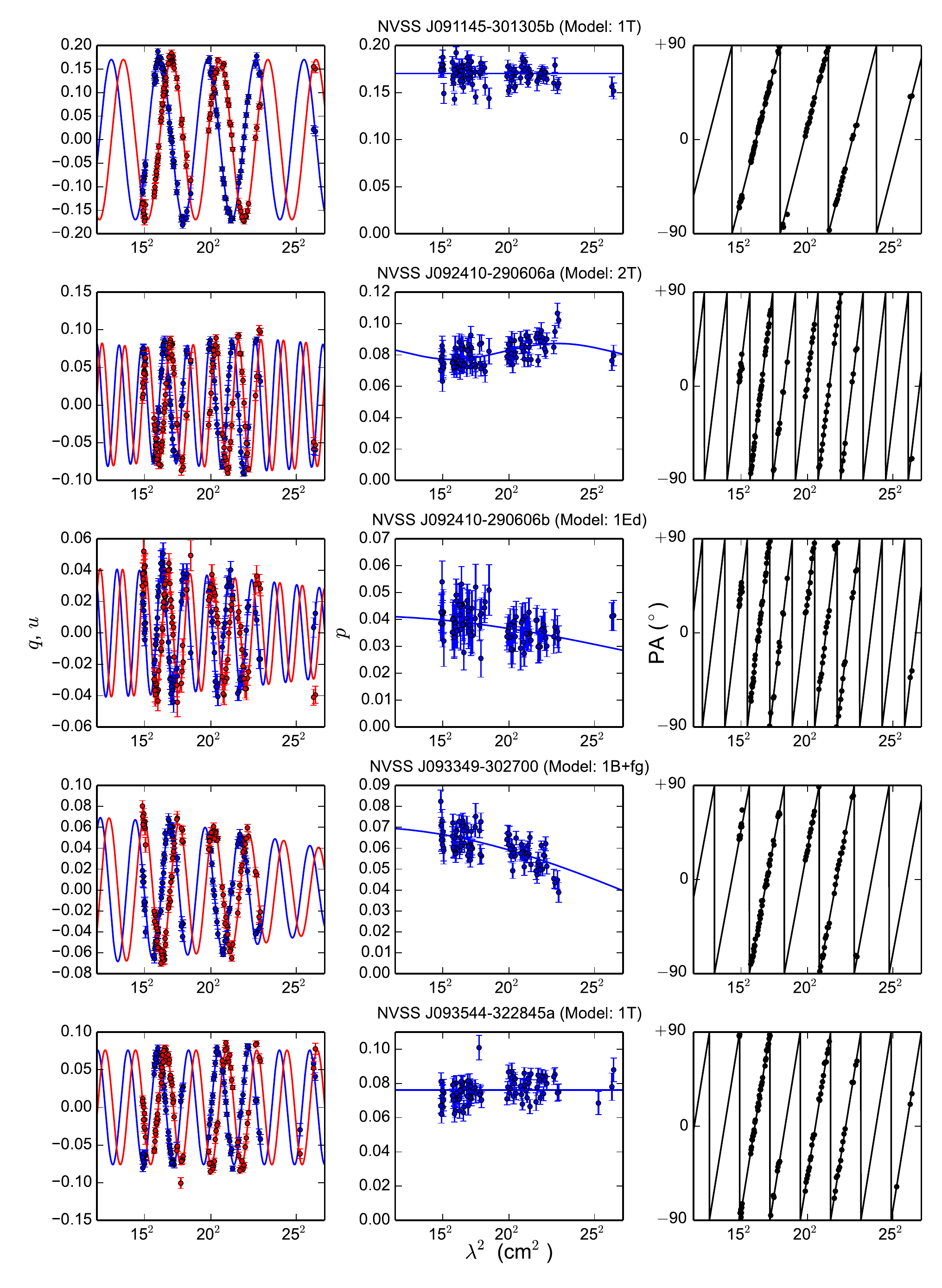}
\caption{(Continued) \textit{QU}-fitting results of our polarised target sources.}
\end{figure*}

\begin{figure*}
\ContinuedFloat
\includegraphics[width=\doublecolumnwidth]{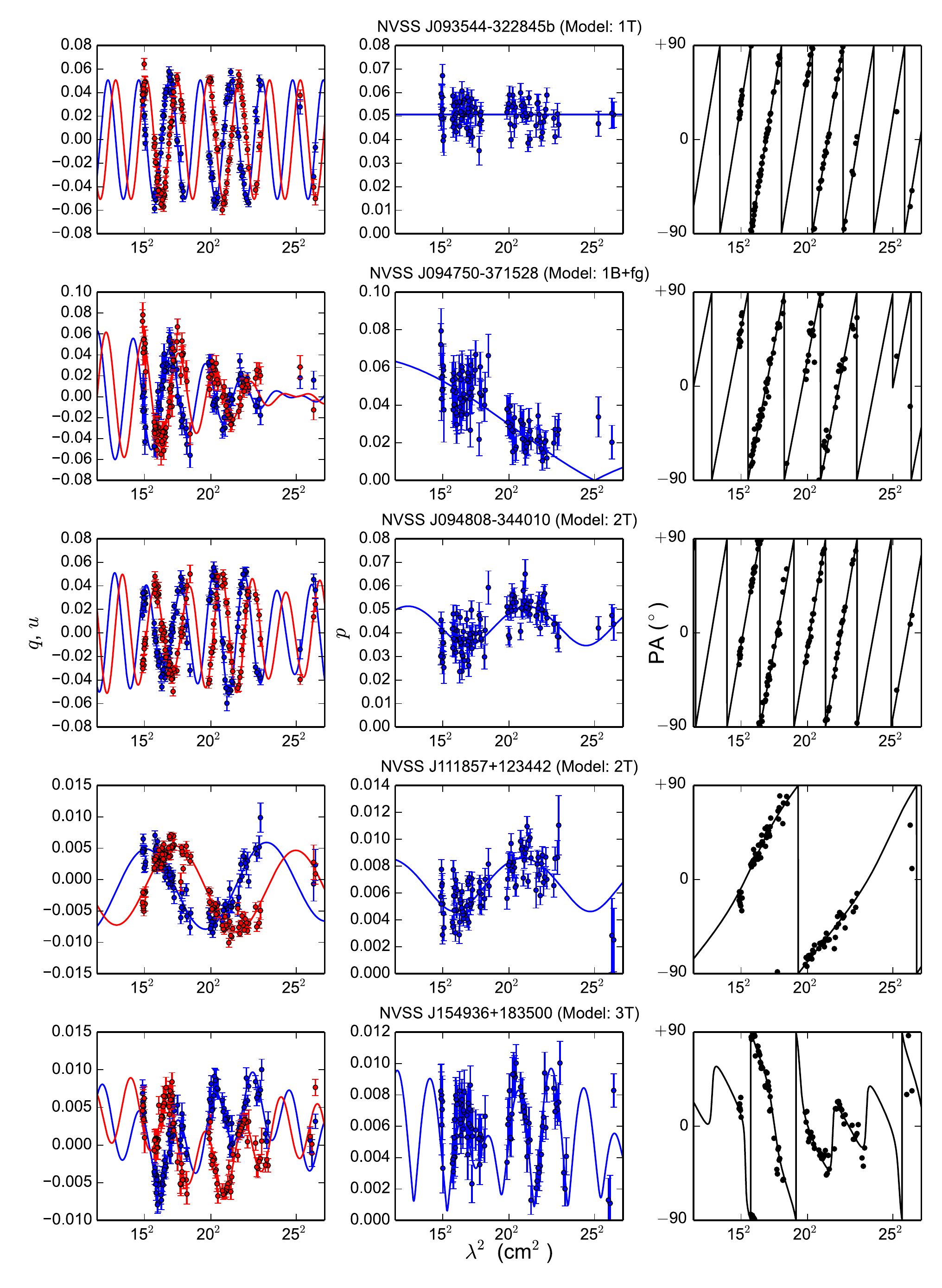}
\caption{(Continued)\textit{QU}-fitting results of our polarised target sources.}
\end{figure*}

\begin{figure*}
\ContinuedFloat
\includegraphics[width=\doublecolumnwidth]{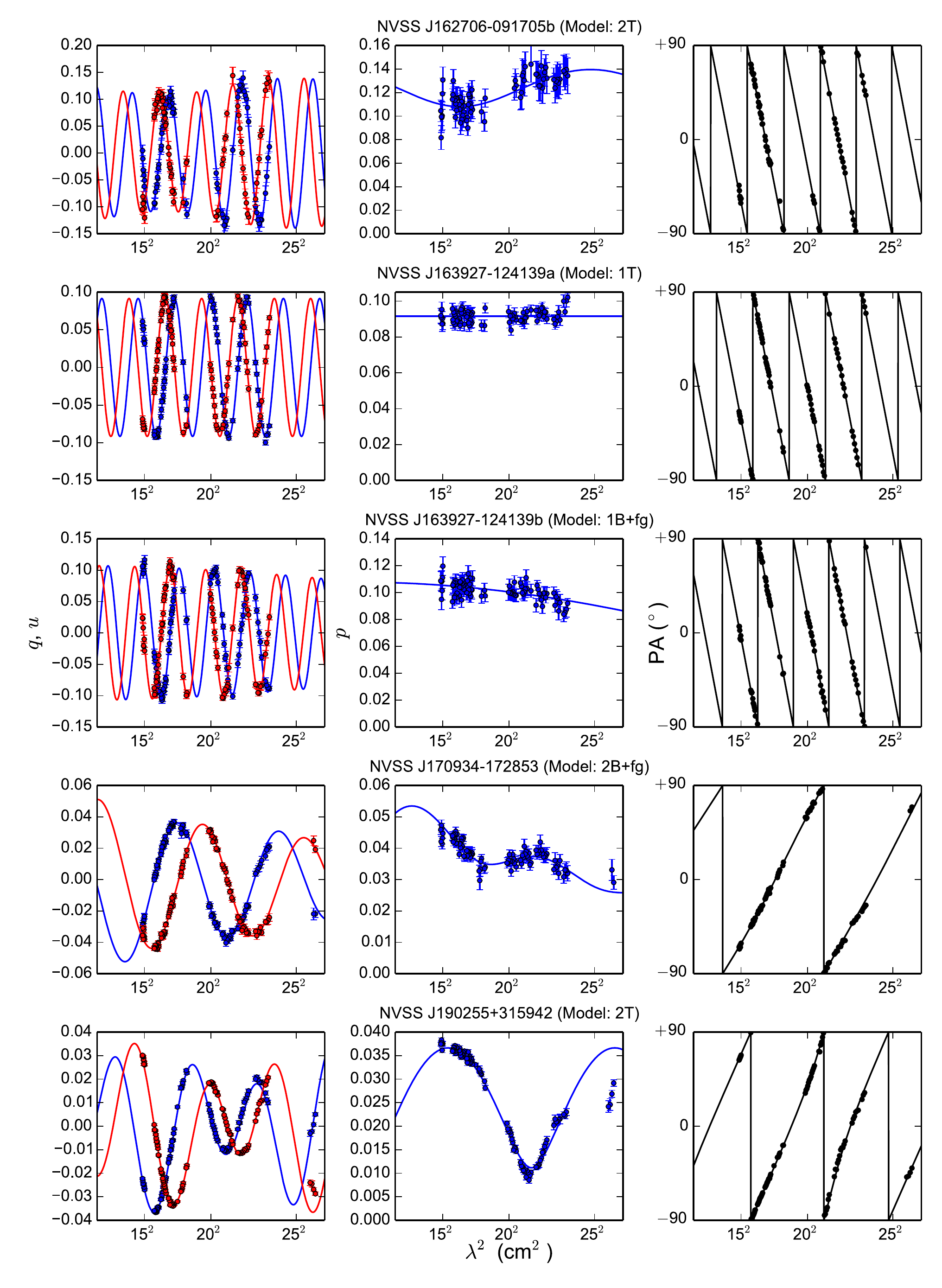}
\caption{(Continued) \textit{QU}-fitting results of our polarised target sources.}
\end{figure*}

\begin{figure*}
\ContinuedFloat
\includegraphics[width=\doublecolumnwidth]{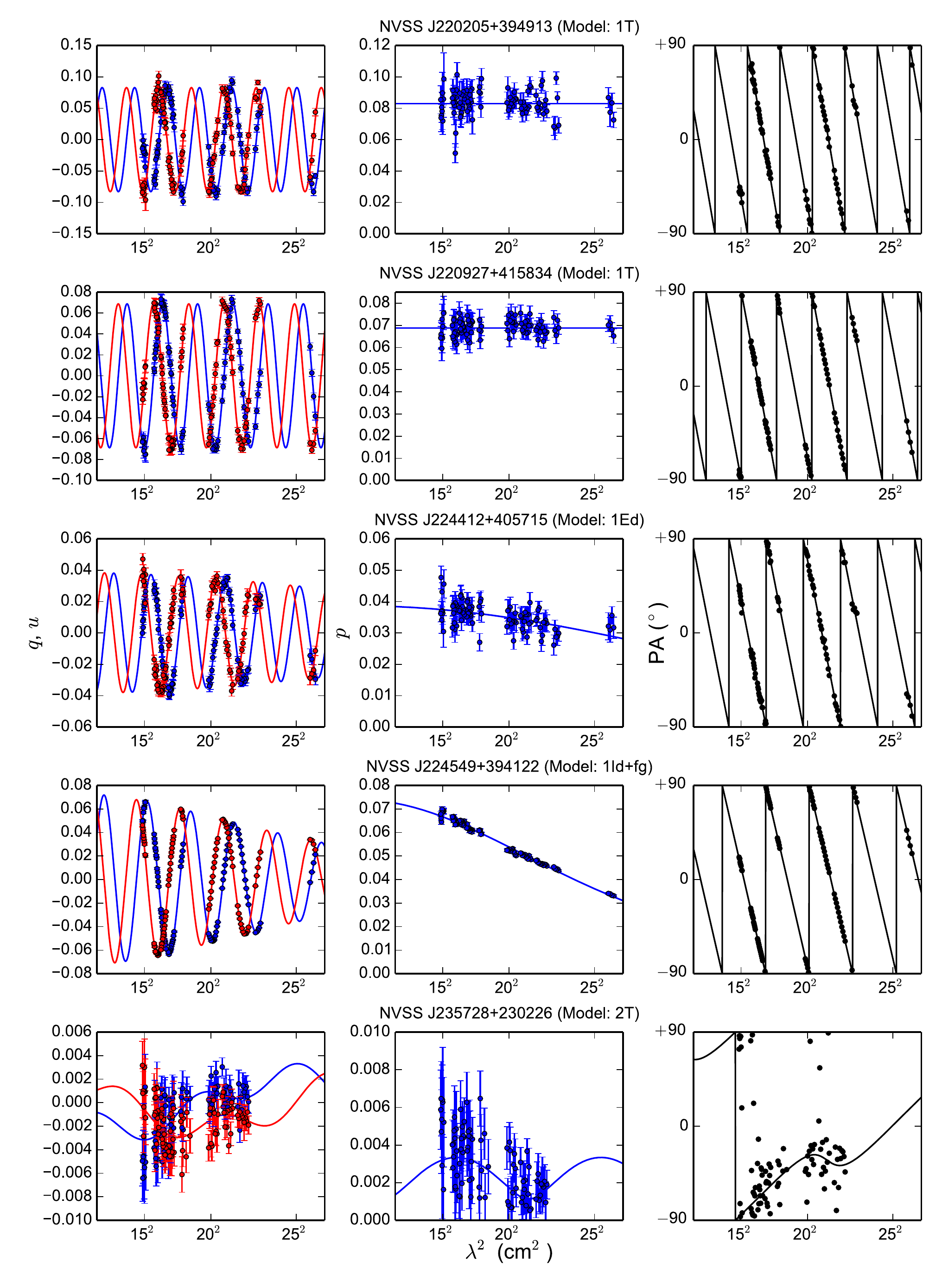}
\caption{(Continued) \textit{QU}-fitting results of our polarised target sources.}
\label{lastpage}
\end{figure*}

\end{document}